\documentclass[%
 reprint,
 superscriptaddress,
 amsmath,amssymb,
 prl,
floatfix,
]{revtex4-2}

\usepackage{graphicx}
\usepackage{dcolumn}
\usepackage{bm}
\usepackage{array}
\newcolumntype{P}[1]{>{\centering\arraybackslash}p{#1}}
\usepackage[hidelinks]{hyperref}
\usepackage{booktabs}
\usepackage{upgreek}
\usepackage{comment}
\usepackage{verbatim}
\usepackage{amssymb}
\usepackage{multirow}
\usepackage{makecell}
\usepackage{aas_macros}

\usepackage[
separate-uncertainty = true,
multi-part-units=single
]{siunitx}
\usepackage{xcolor}
\usepackage[version=4]{mhchem}
\usepackage[normalem]{ulem}
\usepackage{tablefootnote}

\usepackage[section]{placeins} 

\usepackage{lineno}

\newcommand\sonec{S1${c}$}
\newcommand\stwoc{S2${c}$}

\DeclareSIUnit\keVnr{keV_{nr}}
\DeclareSIUnit\keVee{keV_{ee}}

\newcommand\XeOneTwoNine{\ce{^{129}Xe}}

\newcommand\XeOneThreeOne{\ce{^{131}Xe}}

\begin{document}


\title{Searches for Light Dark Matter and Evidence of Coherent Elastic Neutrino-Nucleus Scattering of Solar Neutrinos with the LUX-ZEPLIN (LZ) Experiment} 

\author{D.S.~Akerib}
\affiliation{SLAC National Accelerator Laboratory, Menlo Park, CA 94025-7015, USA}
\affiliation{Kavli Institute for Particle Astrophysics and Cosmology, Stanford University, Stanford, CA  94305-4085 USA}

\author{A.K.~Al Musalhi}
\affiliation{University College London (UCL), Department of Physics and Astronomy, London WC1E 6BT, UK}

\author{F.~Alder}
\affiliation{University College London (UCL), Department of Physics and Astronomy, London WC1E 6BT, UK}

\author{B.J.~Almquist}
\affiliation{Brown University, Department of Physics, Providence, RI 02912-9037, USA}

\author{C.S.~Amarasinghe}
\affiliation{University of California, Santa Barbara, Department of Physics, Santa Barbara, CA 93106-9530, USA}

\author{A.~Ames}
\affiliation{SLAC National Accelerator Laboratory, Menlo Park, CA 94025-7015, USA}
\affiliation{Kavli Institute for Particle Astrophysics and Cosmology, Stanford University, Stanford, CA  94305-4085 USA}

\author{T.J.~Anderson}
\affiliation{SLAC National Accelerator Laboratory, Menlo Park, CA 94025-7015, USA}
\affiliation{Kavli Institute for Particle Astrophysics and Cosmology, Stanford University, Stanford, CA  94305-4085 USA}

\author{N.~Angelides}
\affiliation{University of Zurich, Department of Physics, 8057 Zurich, Switzerland}

\author{H.M.~Ara\'{u}jo}
\affiliation{Imperial College London, Physics Department, Blackett Laboratory, London SW7 2AZ, UK}
\affiliation{STFC Rutherford Appleton Laboratory (RAL), Didcot, OX11 0QX, UK}

\author{J.E.~Armstrong}
\affiliation{University of Maryland, Department of Physics, College Park, MD 20742-4111, USA}

% ...

\collaboration{The LZ Collaboration} 

\begin{abstract}
We present searches for light dark matter (DM) with masses 3--9 GeV/$c^2$ in the presence of coherent elastic neutrino-nucleus scattering (CE$\nu$NS) from $^{8}$B solar neutrinos with the LUX-ZEPLIN experiment. This analysis uses a 5.7~tonne-year exposure with data collected between March 2023 and April 2025. 
In an energy range spanning 1--6~keV, we report no significant excess of events attributable to dark matter nuclear recoils, but we observe a significant signal from $^{8}$B CE$\nu$NS interactions that is consistent with expectation. We set world-leading limits on spin-independent and spin-dependent-neutron DM-nucleon interactions for masses down to 5~GeV/$c^2$. In the no-dark-matter scenario, we observe a signal consistent with $^{8}$B CE$\nu$NS events, corresponding to a $4.5\sigma$ statistical significance. This is the most significant evidence of $^{8}$B CE$\nu$NS interactions and is enabled by robust background modeling and mitigation techniques. This demonstrates LZ's ability to detect rare signals at keV-scale energies.
\end{abstract}

\maketitle

\emph{Introduction}---Weakly interacting massive particles (WIMPs) remain a highly motivated dark matter (DM) candidate in the GeV$/c^{2}$--TeV$/c^{2}$ mass range~\cite{akerib2022snowmass2021,Billard2022}. The LUX-ZEPLIN (LZ) experiment currently sets the strongest constraints on DM-nucleon scattering cross-sections for masses greater than 9~GeV/$c^{2}$~\cite{LZ:2024WIMP}. 
Light DM candidates with masses less than 9~GeV/$c^{2}$ are well-motivated~\cite{Kaplan:2009ag,Cohen:2010kn}, but present significant analysis challenges as they produce lower energy recoils. Additionally, at the increasingly smaller cross-sections probed by current generation xenon DM experiments, the sensitivity to light DM is limited by solar neutrinos undergoing coherent elastic neutrino-nucleus scattering (CE$\nu$NS) which form an irreducible `neutrino-fog' background~\cite{Billard:2013qya,Ruppin:2014bra,OHare:2021utq,akerib2022snowmass2021}. Neutrinos from $^{8}$B decays in the Sun~\cite{Bahcall:2004pz} produce CE$\nu$NS interactions, resulting in a nuclear recoil spectrum in xenon that closely resembles that of a $5.5$~GeV/$c^{2}$ DM candidate. Previous results from the PandaX-4T and XENONnT experiments have revealed indications of solar $^{8}$B CE$\nu$NS interactions at significances lower than 3$\sigma$~\cite{Panda:8B,XenonNT:8B}, and have set limits on light DM nucleon cross-sections in this regime~\cite{XenonNT:LDM-NeutrinoFog,PandaX2025}. In this Letter we present the first nuclear recoil DM search from LZ for DM of masses 3--9~GeV/$c^{2}$ and the first evidence of solar neutrino CE$\nu$NS interactions on xenon, using a 5.7 tonne-year exposure. 

\emph{Experiment}---The LZ experiment~\cite{lztdr,LZExperiment} operates underground in the Sanford Underground Research Facility (SURF) in Lead, South Dakota, USA, with a 4300 meter water-equivalent overburden to shield from cosmic radiation. At the center of the experiment is a dual-phase time projection chamber (TPC) containing seven tonnes of active liquid xenon. The TPC is surrounded by two subsystems, the liquid xenon Skin and the Outer Detector (OD), which serve as active vetoes to reject $\gamma$-ray and neutron backgrounds. The entire apparatus is immersed in 230 tonnes of ultrapure water, which passively shields from radioactivity in the underground cavern. 

Particle interactions in the liquid xenon target typically produce either electron recoils (ERs) or nuclear recoils (NRs), generating prompt vacuum ultraviolet scintillation (S1) and ionization. Ionization electrons drift upward under the influence of an electric field. They are extracted into a thin layer of gaseous xenon above the liquid surface where they produce a delayed electroluminescence signal (S2). The S1 and S2 photons are observed by photomultiplier tube arrays above and below the active liquid xenon volume. The time interval between the S1 and the S2 pulses measures the drift time of the ionization, which is proportional to the interaction depth ($z$), while the transverse position $(x,y)$ is reconstructed from the hit pattern of the S2 light in the top PMT array. 

\emph{Dataset}---The dataset for this analysis (WS2025) includes 546~live~days collected between March 2023 and April 2025, extending the WS2024 dataset used in our previous dark matter search~\cite{LZ:2024WIMP}. Dedicated americium-beryllium (AmBe) and deuterium-deuterium (DD) neutron calibrations were performed during this period. The TPC drift field was established at 97~V/cm and the extraction field in the liquid phase was 3.4~kV/cm.

\begin{figure}[!t]
	\centering
	\includegraphics[width=1 \columnwidth]{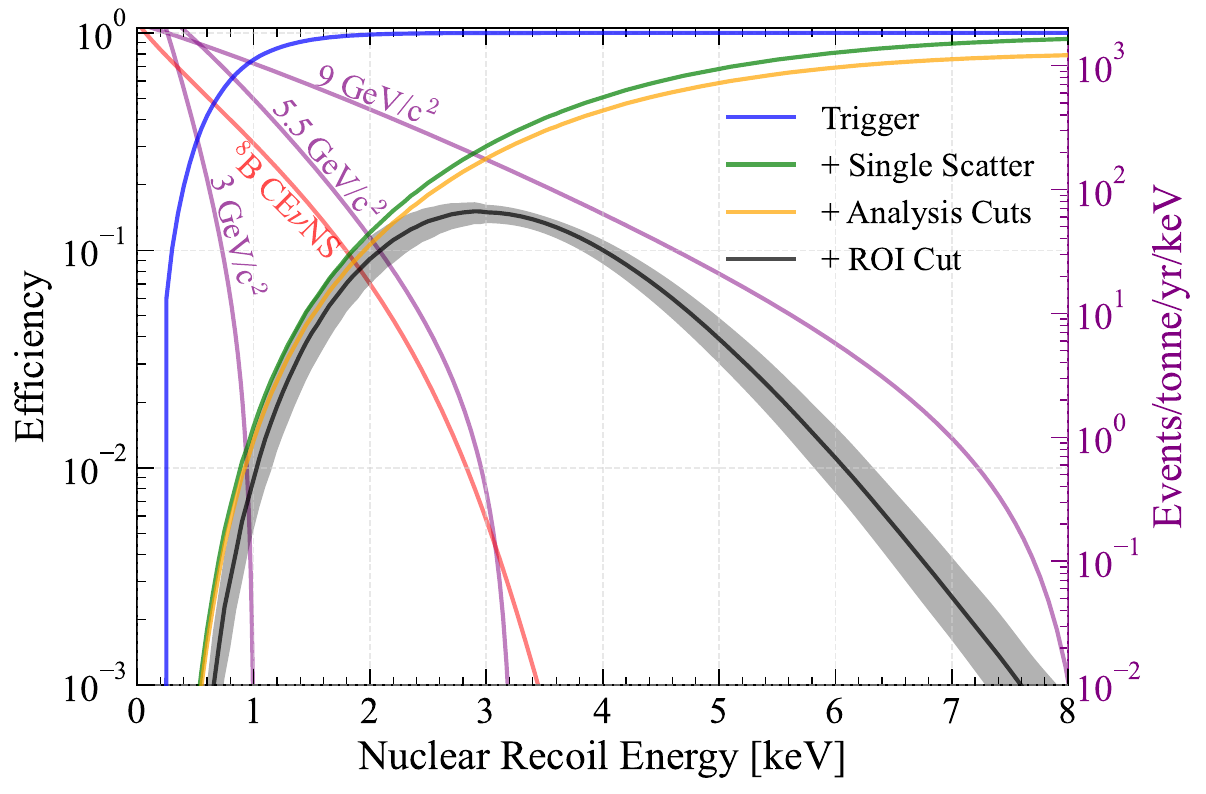}
	\caption{Detection efficiency as a function of NR energy, after applying the S2 trigger (blue), $\geq3$-fold SS reconstruction (green), analysis selections (orange), and region-of-interest (black) in sequence. The uncertainty band (gray) includes the uncertainties associated with the yields and fluctuations in the NR response model and the efficiency of single-scatter reconstruction and analysis selections. Also shown are the DM recoil spectra (purple) for masses of 3~GeV/$c^{2}$, 5.5~GeV/$c^{2}$, and 9~GeV/$c^{2}$, assuming a cross-section of 10$^{-44}$~cm$^{2}$, before the application of detection efficiency. The predicted $^{8}$B CE$\nu$NS recoil spectrum (red) is similar to that of a DM particle with mass 5.5~GeV/$c^{2}$.}
	\label{fig:NRDetEff}
\end{figure}

We collect data using the data acquisition system described in Ref.~\cite{LZ_DAQ}, with a trigger efficiency of 95~$\pm$~1\% for S2 sizes corresponding to 4 extracted electrons. An event reconstruction algorithm identifies single scatter (SS) events with one prominent S1 and S2, with the efficiency shown in Figure~\ref{fig:NRDetEff}. The loss in efficiency at low energy is driven by requiring that the S1 is observed in at least three PMTs. We quantify these inefficiencies with waveform simulations and calibration datasets:  beta decays from an injected tritiated methane (CH$_{3}$T) source and AmBe neutron calibrations. Pulse shape variations from photon arrival times and electron drift diffusion result in additional losses, and are accounted for in the detector response modeling.

The calibration program of the TPC~\cite{LZ_calibrations} allows for monitoring of detector spatial and temporal uniformity. The $(x,y,z)$ position- and time-dependent responses of the S1 and S2 signals are corrected using $^{222}$Rn and $^{218}$Po $\alpha$-decay events present throughout the detector as in~\cite{LZ:2024WIMP}, allowing the definition of corrected quantities $\text{S}1c$ and $\text{S}2c$ in units of photons detected (phd). 
The scintillation and ionization gains for quanta produced in the liquid xenon are $g_1=0.112\pm0.002$~phd/photon and $g_2=34.0\pm0.9$~phd/electron~\cite{LZ:2024WIMP}.

The region-of-interest (ROI) is restricted to SS events where the S1 is observed in at least three PMTs with the S1$c$ size that lies between 2 and 15~phd. The search parameter space spans an S2 range of 3.5--14.5 electrons (44.5~phd per extracted electron), disjointed from the ROI of~\cite{LZ:2024WIMP}. The ROI approximately corresponds to nuclear recoil energies from 1--6~keV.

\emph{Selections}---This ROI is contaminated by events from delayed charge and light emission after large energy depositions. After such events, we reject data for a period that scales with the size of the energy deposition, incurring a 16\% loss of exposure. The median vetoing pulse area is 10$^{5.8}$~phd, and the mean vetoed time is 44~ms. After removing periods of unstable detector conditions and elevated photon or electron rates, 417 live days remain. 

A fiducial volume (FV) is defined to reject particle interactions occurring in proximity to the stainless steel high voltage grids and TPC wall. S2s in this analysis suffer from worsened $(x,y)$ position resolution~\cite{LZ-backgrounds}, compared to the larger S2s used in Ref.~\cite{LZ:2024WIMP}. This motivates a stricter radial FV boundary, on average 5.4~cm from the wall and 15~cm from the PTFE-embedded high voltage resistors. We require $<0.01$ total events originating from the wall which sets the definition of the radial FV boundaries as a function of depth and azimuth. The vertical limits are unchanged from~\cite{LZ:2024WIMP}. This results in a FV mass of 5.09~$\pm$~0.15 tonnes compared to 5.5~$\pm$~0.2 in~\cite{LZ:2024WIMP}, which is determined using tritium events. Events with S2s consistent with localized grid emission~\cite{Akerib:2025rzt} are also removed through time and S2 $(x,y)$ position-based selections using a computer vision algorithm~\cite{opencv}, resulting in an additional 1\% loss of exposure. The remaining 5.7~$\pm$~0.2~tonne-years of exposure is used in this analysis.

Veto coincidence requirements reject external backgrounds such as radiogenic neutrons from spontaneous fission or ($\alpha$,n) reactions~\cite{LZ:2024WIMP}. Prompt coincidences of TPC events with pulses in the OD ($\leq$~300~ns) or Skin ($\leq$~250~ns) reject neutron scatters. Delayed coincidences (using a $\leq$~600~$\mu$s time window following the main S1) target neutrons which thermalize, capture in the gadolinium-loaded liquid scintillator, and produce cascades of $\gamma$-rays following nuclear de-excitation. AmBe data taken during this period allowed for measuring a delayed neutron tagging efficiency in good agreement with measurements from americium-lithium (AmLi) data in~\cite{LZ:2024WIMP}. Corresponding neutron simulations, corrected using AmBe and AmLi calibration data, predict the delayed (delayed + prompt) tagging efficiency of radiogenic neutrons to be 87~$\pm$~2~\% (92~$\pm$~4~\%)~\cite{LZ:2024WIMP}. 

Finally, a series of selection criteria remove events inconsistent with xenon recoil signals in the fiducial volume. These criteria exploit the expected S1 and S2 spatial pattern and pulse shapes of genuine events. Each selection was initially defined to maintain high acceptance of signals from calibration data. Selections with simple functional forms were then refined utilizing a covariance matrix adaptation evolution strategy (CMA-ES)~\cite{hansen2019pycma} which simultaneously optimizes the signal-to-background ratio across multiple dimensions and outputs selections defined in the observables provided to the algorithm. The efficiency loss from these selections is shown in Figure~\ref{fig:NRDetEff}.

\emph{Bias Mitigation}---To reduce analyzer bias, we fabricate and insert artificial signal-like events into the raw dataset~\cite{Klein_and_Roodman_2005} by combining real S1 and S2 pulses from sequestered CH$_{3}$T and AmLi calibration data, respectively. We create two types of events: high mass DM-like scatters, following an exponential plus flat energy spectrum, described in Ref.~\cite{LZ:2024WIMP}; and, for the latter 59\% of this dataset, we additionally inject $^{8}$B CE$\nu$NS-like scatters. The former type is designed for the $\geq$~9~GeV/$c^2$ DM search and has a small expected contribution in this ROI. The artificial scatters consist of detectable S1 and S2 pulses, with the injected number of events after all selection criteria drawn from a Poisson distribution of mean 1.75. The true number and type of artificial events inserted into the dataset was hidden from data analyzers until after the analysis was finalized. Subsequently, it was revealed that only one artificial single event was injected into the data within our ROI and fiducial volume selection. This event also passed all our data selection criteria. While prior bias mitigation efforts using similar techniques injected substantially more events~\cite{Savage_1986, Sanchez_2003, Klein_and_Roodman_2005}, the goal of this injection is not to mask the presence of $^{8}$B interactions, but to mask their rate.  The mean injection rate was intended to be comparable to 1$\sigma$ fluctuations on the expected $^{8}$B signal in the latter 59\% of this dataset. Additionally, the expected rate of $^{8}$B CE$\nu$NS events for injection was calculated prior to the results from a new dedicated low-energy calibration campaign and finalization of cuts and efficiencies; this means the mean of the Poisson distribution (1.75) was smaller than the square-root of the final expected rate reported here. See the Supplemental Materials for more information on bias mitigation.

\emph{Response Modeling}---The detector response model used to predict the response of xenon nuclear recoils is based on \textsc{nest}~2.4.5~beta~\cite{nest_2_4_5,szydagis2022review} and tuned using a dedicated DD neutron dataset. The LZ DD generator provides a well-characterized source of nearly-monoenergetic 2.45~MeV neutrons producing NRs from the smallest detectable signals to the endpoint recoil, 1--74~keV~\cite{LUX_DD,LUX:2022qxb}. 

We constrain the \textsc{nest} model describing the charge and light yields ($Q_y$ and $L_y$, respectively) and their fluctuations as a function of energy using a Markov Chain Monte Carlo-based fitting method implemented with the \textsc{emcee} package~\cite{emcee}. Neutron recoils out to the DD endpoint are included in this fit  to constrain the \textsc{nest} model parameters across a large energy range, including higher energies outside of the ROI of this analysis. The DD dataset has limited ability to constrain the low-energy-specific yield parameters ($\lesssim2.5$~keV), motivating our choice to enforce \textsc{nest} priors derived from other existing measurements~\cite{LUX_DD,LUX:2022qxb,Lenardo:2019fcn} for these parameters.

We include two important modifications to the \textsc{nest} model described in Ref.~\cite{szydagis2022review}. First, fluctuations in S1 and S2 sizes are allowed to vary via the Fano-like factors $F_{ex}$ and $F_{i}$, which control the variances of excitons and ions, respectively. We model $F_{ex}$ as a linear function of exciton number where the slope and intercept are allowed to float freely. We find that $F_{i}$ can be modeled as a constant. Second, we introduce an additional parameter to allow recombination fluctuations, or the fluctuations governing the trade-off between $Q_y$ and $L_y$, to be either sub- or supra-binomial. Both modifications are needed to achieve a good fit across the DD energy spectrum. Further details of this fitting procedure, including best fit parameters and corresponding yields models, are discussed in the Supplemental Materials and an upcoming publication.

\begin{table*}[!t]
  \centering
    \caption{Expected and best-fit counts for each background and/or signal component in various fit scenarios to the 19 observed events, as well as the value of the Gaussian-constrained nuisance parameter $\sigma_\text{eff}$. The second column shows the predicted counts and  $\sigma_\text{eff}$ value, the latter of which is 0~$\pm$~1 by construction. The next three columns show fits where the $^{8}$B signal is incorporated as a background component following the expectation in the first column; results from a background-only fit, where the $^{8}$B signal is constrained, and signal-plus-background fits assuming a spin-independent dark matter particle with mass 3~GeV/$c^{2}$ and 9~GeV/$c^{2}$ are shown. The last column shows the fit with the $^{8}$B rate unconstrained in the no-dark-matter hypothesis. The spin-independent (SI) DM, neutron, and $^{8}$B CE$\nu$NS uncertainties include systematics associated with the NR response modeling, the detection uncertainties, and the flux uncertainty ($^{8}$B CE$\nu$NS only). The fitted values of $\sigma_\text{eff}$ represent the shifts in the fitted component rates with respect to their predicted values due to the combined NR response modeling and detection efficiency uncertainties, in units of standard deviations.}
  \begin{tabular}{P{2.8cm} || P{2.4cm} | P{2.9cm}  P{2.7cm} P{2.7cm} | P{3.0cm} }
  \hline \hline 
  \textbf{Components} & 
  \begin{tabular}{@{}c@{}}\textbf{Expectation}\end{tabular}& 
  \begin{tabular}{@{}c@{}}\textbf{Background-Only} \\ \textbf{Fit}\end{tabular} &
  \begin{tabular}{@{}c@{}}\textbf{3~GeV/$c^{2}$} \\ \textbf{Fit}\end{tabular} &
  \begin{tabular}{@{}c@{}}\textbf{9~GeV/$c^{2}$} \\ \textbf{Fit}\end{tabular} 
  & \begin{tabular}{@{}c@{}}\textbf{$^{8}$B CE$\nu$NS} \\ \textbf{Unconstrained Fit}\end{tabular} \\
  \hline
    SI DM & - & - & $0.4\substack{+5.4 \\ -0.4}$ & $0.0\substack{+4.0 \\ -0.0}$ & - \\
    $^{8}$B CE$\nu$NS & $20.6\substack{+8.9 \\ -6.8}$ & $15.0\substack{+4.0 \\ -3.4}$ & $14.7\substack{+4.2 \\ -3.9}$ & $15.0\substack{+3.9 \\ -3.4}$ & $12.3\substack{+7.0 \\ -5.4}$ \\
    Accidental coinc. & $6.6~\pm~0.3$ & $6.5~\pm~0.3$ & $6.5~\pm~0.3$ & $6.5~\pm~0.3$ & $6.6~\pm~0.3$ \\
    Detector neutrons & $0.04\substack{+0.25 \\ -0.04}$  & $0.1\substack{+0.2 \\ -0.1}$ & $0.1\substack{+0.2 \\ -0.1}$ & $0.1\substack{+0.2 \\ -0.1}$ & $0.1\substack{+0.2 \\ -0.1}$\\
    \hline
    Total & $27.2\substack{+10.1 \\ -7.3}$ & $21.6\substack{+4.7 \\ -3.8}$ & $21.7\substack{+6.2 \\ -2.8}$ & $21.6\substack{+5.0 \\ -2.5}$ & $18.9\substack{+7.0 \\ -5.5}$ \\
    \hline
    $\sigma_\text{eff}$ & $0~\pm1~$ & $-0.81\substack{+0.59 \\ -0.60}$ & $-0.86\substack{+0.63 \\ -0.71}$ & $-0.81\substack{+0.58 \\ -0.60}$ & $0~\pm1~$\\
    \hline \hline
  \end{tabular}
  \label{tab:fitResults}
\end{table*}

\emph{Backgrounds}---Accidental coincidence events and neutrons are background components in the search for $^{8}$B CE$\nu$NS interactions. Events caused by electrons and $\gamma$-ray backgrounds produce S2 signals above the ROI of this analysis and are thus negligible. For the dark matter search, $^{8}$B CE$\nu$NS is considered a background and is discussed in the following section.

Accidental coincidence events are comprised of S1-S2 pairs where the pulses do not originate from a common energy deposition. Instrumental effects, such as PMT dark counts~\cite{Akerib:2021pfd} and spurious electron emission from the high voltage electrodes~\cite{Akerib:2025rzt}, produce isolated S1 and S2 pulses which can pile up within an event window. These events are modeled using a data-driven technique in which isolated pulses are combined at the waveform level to form synthetic SS events. In this ROI, the spectra of isolated S1 and S2 pulses are sensitive to variations in electron and photon rates occurring on millisecond timescales. Thus we perform an environment-matched pairing procedure where we pair isolated S1 and S2 pulses according to the ambient electron and photon rates. The result is a high-fidelity description of real accidental events.

A pure population of accidental coincidences can be found by searching for events with unphysical drift times (UDT), where the apparent drift time exceeds that of events originating at the cathode ($\gtrsim$~1~ms). This population provides a data-driven constraint on the accidental coincidence rate and is used to determine the normalization for the accidental model. The model is extensively validated with several ancillary datasets which probe a range of environments: UDT events, events with elevated photon and electron rates, AmLi calibration data, and events that fail the S1 and S2 pulse-based selections. The resulting goodness-of-fit tests~\cite{BAKER1984437} comparing the data and model prediction in binned S1$c$, S2$c$, and \{\sonec,~\stwoc\} distributions in these ancillary datasets all pass at a significance level of 0.05. More details are discussed in the Supplemental Materials and an upcoming publication.  

After all selection criteria, the predicted number of accidental coincidences is 6.6~$\pm$~0.3 events in the ROI for the 5.7~tonne-year exposure. The uncertainty is driven by both the number of UDT events at the point of normalization and the number of unique S1 and S2 pulses used in the accidental model. Additional systematic uncertainties are deemed unnecessary based on the good data-model agreement observed across the ancillary datasets. These datasets validate the normalization procedure and confirm the ability to accurately model populations with varying underlying accidental coincidence sources.

\begin{figure}[!t]
	\centering
	\includegraphics[width=1 \columnwidth]{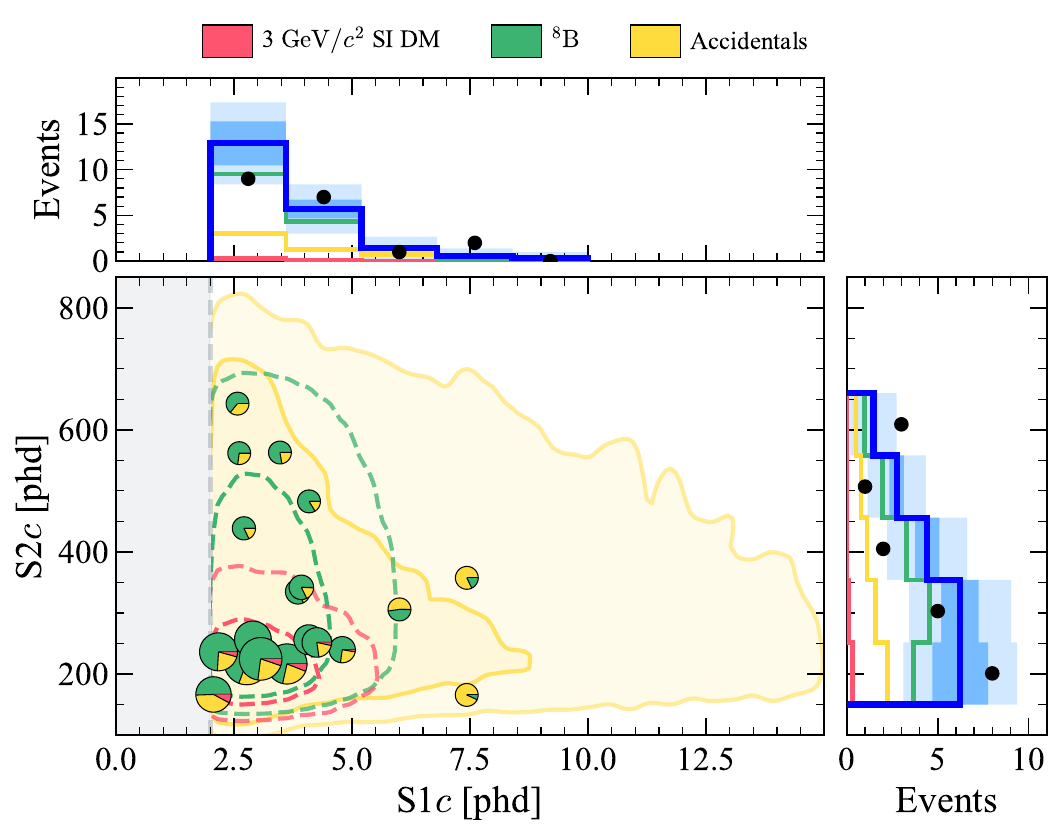}
	\caption{The 19 events comprising the final dataset passing all selections are shown as individual pie charts in the observable space \{\sonec,~\stwoc\}. Each pie chart depicts the relative fit contributions from the accidental coincidences (yellow) and $^{8}$B CE$\nu$NS (green) background components along with a 3~GeV/$c^{2}$ DM candidate (pink). The relative size of each pie chart is proportional to its contribution to 3~GeV/$c^{2}$ DM events. Contours enclose $1\sigma$ and $2\sigma$ model distributions. The data (black points) and best-fit model (solid blue line) in \sonec~and \stwoc~projections are also shown. Shaded light blue (dark blue) bands depict the total (systematic-only) model uncertainty. The detector neutron component is subdominant and therefore not shown. }
	\label{fig:DataModel}
\end{figure}

\emph{Dark Matter Search}---After the application of all data selection criteria and the removal of the one artificial event, 19 events remain in the ROI (see Figure~\ref{fig:DataModel}). The search for dark matter is conducted with a two-sided unbinned profile likelihood ratio test statistic defined in the observable space \{\sonec,~\stwoc\}~\cite{cowan2011asymptotic,baxter2021recommended}.

The components of the likelihood describing particle interactions --- detector neutrons, $^{8}$B CE$\nu$NS, and the dark matter signal --- are constructed with the xenon response model described above and an event simulation framework~\cite{akerib2021simulations}. We account for the systematic uncertainty on the light and charge yields, the quanta fluctuations, and the detector efficiency in the likelihood. This is incorporated through a single Gaussian constraint with nuisance parameter $\sigma_\text{eff}$ controlling the combined rate uncertainty, the scale of which depends on the individual recoil spectra. The change in rate due to a +1 ($-$1) standard deviation shift in $\sigma_\text{eff}$ varies as a function of DM mass, from  $+$81\% ($-$60\%) at 3~GeV/$c^{2}$ to $+$21\% ($-$20\%) at 9~GeV/$c^{2}$.

We constrain the detector neutron rate \textit{in situ} by including an additional dataset in the likelihood which requires events to pass all data selection criteria but fail the delayed coincidence veto. There is one event in this dataset. The best-fit number of neutron events in this sideband dataset is $0.3\substack{+1.7 \\ -0.3}$, which is consistent with predictions from simulations. The neutron veto efficiency, incorporated as a Gaussian constraint, relates this sideband dataset to the main science dataset, while the other background components are related via the false veto probability from random coincidences (3\%). 

The expected number of events in the main science dataset from $^{8}$B CE$\nu$NS interactions, accidental coincidences, and detector neutrons is shown in Table~\ref{tab:fitResults}. The uncertainties are incorporated in the likelihood as constraints on their respective rates.  The accidental rate is Gaussian-constrained. The $^{8}$B CE$\nu$NS rate has an additional 4\% flux uncertainty imposed as a Gaussian constraint~\cite{baxter2021recommended}. 

The best-fit model and data in the \{\sonec,~\stwoc\} analysis space and its projections are also shown in Figure~\ref{fig:DataModel}. The background-only model is evaluated using binned goodness-of-fit tests, with p-values of 0.51, 0.44, and 0.53 in $\text{S}1c$,~$\text{S}2c$, and \{\sonec,~\stwoc\} spaces, respectively. All data-model comparisons show good agreement. 

\begin{figure}[!t]
	\centering
	\includegraphics[width=1 \columnwidth]{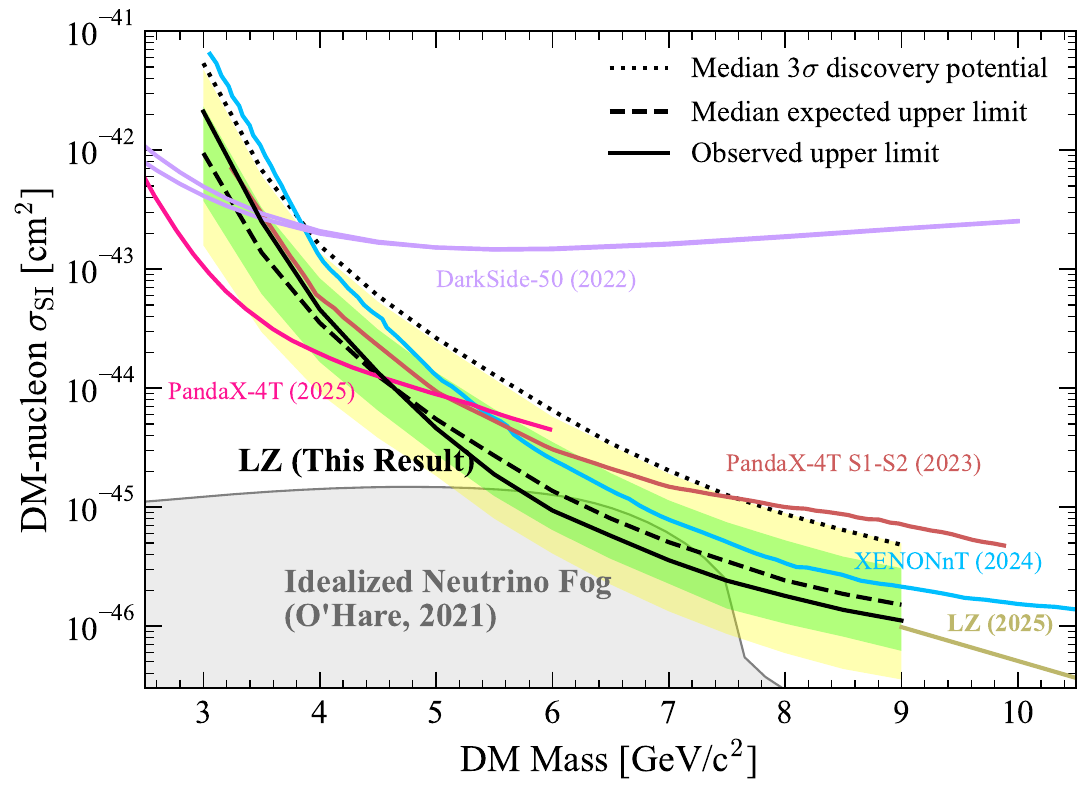}
	\caption{Upper limits (90\% C.L.) on the spin-independent DM-nucleon cross section as a function of DM mass using the 5.7~tonne-year dataset are shown in the solid black line. The range of expected upper limits from background-only experiments are shown as green and yellow regions for 68\% and 95\% of experiments, respectively. The dashed black line shows the median expected upper limit, while the dotted black line shows the median $3\sigma$ discovery potential, using the post-fit model. The onset of the neutrino fog, as defined in~\cite{OHare:2021utq}, is shown as the shaded gray region. Other experimental limits are also shown \cite{XenonNT:LDM-NeutrinoFog,PandaX:2022aac,DarkSide-50:2022qzh,PandaX2025}.}
	\label{fig:WIMPSI}
\end{figure}

The 90\% confidence level upper limit on the spin-independent DM-nucleon cross section as a function of mass is shown in Figure~\ref{fig:WIMPSI}. The observed upper limit spans from 2.1$\times10^{-42}$cm$^{2}$ at 3~GeV/$c^2$ to 1.1$\times10^{-46}$cm$^{2}$ at 9~GeV/$c^2$. Results for the spin-dependent DM-nucleon coupling scenario are shown in the Appendix. 

\emph{Boron-8 CE$\nu$NS Measurement}---We also perform a measurement of the $^{8}$B CE$\nu$NS process on xenon nuclei with the same dataset assuming zero dark matter events. Using the $q_{0}$ test statistic~\cite{cowan2011asymptotic}, we reject the hypothesis of zero $^{8}$B CE$\nu$NS events with a statistical significance of $4.5\sigma$. This significance is driven by our observation of 19 events, which represents a substantial deviation from the total expected backgrounds from accidental coincidences and neutrons (6.6~$\pm$~0.3). The median expected significance is $6.7^{+2.4}_{-2.2}\sigma$. The best-fit value of $^{8}$B CE$\nu$NS counts of $12.3\substack{+7.0 \\ -5.4}$ is shown in the last column of Table~\ref{tab:fitResults} and is consistent with the prediction of $20.6\substack{+8.9 \\ -6.8}$.

\begin{figure}[!t]
	\centering
    \includegraphics[width=1\columnwidth]{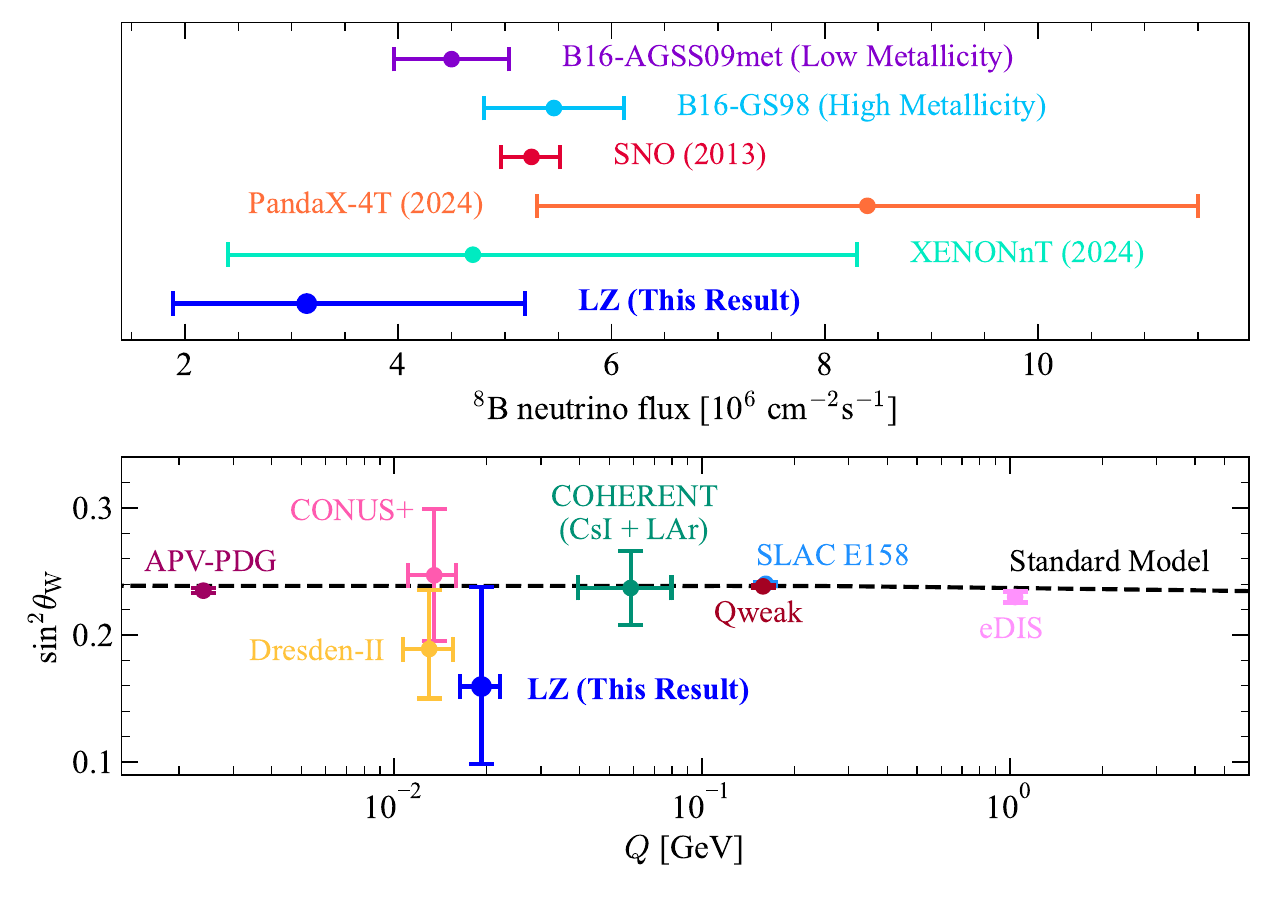}
	\caption{\emph{Top:} Measurement of $^{8}$B solar neutrino flux assuming the SM-predicted CE$\nu$NS cross-section and weak mixing angle. Predictions from low- and high-metallicity solar models are shown~\cite{Vinyoles:2016djt} alongside other experimental measurements \cite{aharmin2013combined,XenonNT:8B,Panda:8B}.
    \emph{Bottom:} Measurement of the weak mixing angle assuming the SNO measurement of $^{8}$B solar-$\nu$ flux and SM CE$\nu$NS cross-section is shown. The median and $\pm1\sigma$ values are depicted in both the top and bottom plots. Also shown are other experimental results~\cite{Wood_1997_APV,Anthony_2005_SLAC_E158,Androic_2018_Qweak,DeRomeri_2023_COHERENT,Sierra_2022_DresdenII,Alpizar:2025wor,DeRomeri_2025_CONUS,Prescott_1979_eDIS_Old} and the SM prediction~\cite{ParticleDataGroup:2024cfk}.}
	\label{fig:8B}
\end{figure}

The detection of the $^{8}$B solar neutrino CE$\nu$NS process enables measurements of various quantities of physical interest (Figure~\ref{fig:8B}) within the Standard Model (SM). The $^{8}$B CE$\nu$NS event rate is dependent on the CE$\nu$NS differential cross-section~\cite{Freedman_1974, Freedman_1977} and the $^{8}$B solar neutrino flux~\cite{baxter2021recommended}. We also include electroweak radiative corrections to the tree-level CE$\nu$NS cross-section following Refs.~\cite{Erler_2013, Marciano_1980,rjamesthesis}. Similar calculations are also discussed in Ref.~\cite{Tomalak_2021}. In deriving our $^{8}$B CE$\nu$NS signal, we assume the neutrino energy spectrum from Ref.~\cite{Bahcall_1996}. Employing an alternative $^{8}$B neutrino spectra from Ref.~\cite{Longfellow_2023} shifts the predicted $^{8}$B counts by less than 6\%, which is subdominant to our systematic uncertainty. If we assume the SNO measured $^{8}$B neutrino flux~\cite{aharmin2013combined}, we can directly probe average CE$\nu$NS cross-section on xenon and specifically the weak-mixing angle~\cite{Maity:2024aji}. The $^{8}$B solar neutrino flux is found to be $3.1\substack{+2.1 \\ -1.3}\times10^{6}$~cm$^{-2}$s$^{-1}$, in good agreement with Ref.~\cite{Vinyoles:2016djt}. The flux-weighted CE$\nu$NS cross-section on xenon is found to be $7.4\substack{+4.8 \\ -3.0}\times10^{-40}$~cm$^{2}$. The SM weak mixing angle at an average momentum transfer of 19.2~$\pm$~2.9~MeV/$c$ is found to be $\sin^{2}\theta_{W}=0.16\substack{+0.08 \\ -0.06}$. We compute the momentum transfer as $Q=\sqrt{2m_{\mathrm{Xe}}E_{r}}$, averaging over the natural isotopic abundances of xenon to calculate the xenon nuclear mass $m_{\mathrm{Xe}}$, and using the mean $^{8}$B CE$\nu$NS nuclear recoil energy $E_{r}$ predicted by our simulation including detection efficiencies (1.51~$\pm$~0.46~keV). All of these measurements are consistent with SM expectations and previous measurements.

\emph{Conclusions}---LZ has achieved limits on SI DM-nucleon interactions that are world-leading above 5~GeV/$c^{2}$. The sensitivity and robustness of this analysis is enabled by careful calibration of the detector nuclear recoil response, with a comprehensive treatment of light and charge yields as well as their fluctuations, and excellent mitigation and modeling of accidental coincidences down to S2s of 3.5~electrons that would otherwise impede DM searches with masses less than 9~GeV/$c^{2}$. We present the strongest evidence to date for CE$\nu$NS on xenon from solar neutrinos at a significance of $4.5\sigma$, which demonstrates LZ's ability to detect nuclear recoil signals of astrophysical origin. Using the detected $^{8}$B CE$\nu$NS interactions, we present measurements of the weak mixing angle and CE$\nu$NS cross-section on xenon. The measured $^{8}$B neutrino flux is found to be consistent with that measured by the SNO collaboration. The LZ experiment continues to take data towards a target 1000-day exposure that will allow for further dark matter searches into the neutrino fog and enable improved measurements of the CE$\nu$NS process.

\emph{Acknowledgements} --- The research supporting this work took place in part at the Sanford Underground Research Facility (SURF) in Lead, South Dakota. Funding for this work is supported by the U.S. Department of Energy, Office of Science, Office of High Energy Physics under Contract Numbers DE-AC02-05CH11231, DE-SC0020216, DE-SC0012704, DE-SC0010010, DE-AC02-07CH11359, DE-SC0015910, DE-SC0014223, DE-SC0010813, DE-SC0009999, DE-NA0003180, DE-SC0011702, DE-SC0010072, DE-SC0006605, DE-SC0008475, DE-SC0019193, DE-FG02-10ER46709, UW PRJ82AJ, DE-SC0013542, DE-AC02-76SF00515, DE-SC0018982, DE-SC0019066, DE-SC0015535, DE-SC0019319, DE-SC0025629, DE-SC0024114, DE-AC52-07NA27344, \& DE-SC0012447. This research was also supported by the UKRI’s Science \& Technology Facilities Council under award numbers ST/W000490/1, ST/W000482/1, ST/W000636/1, ST/W000466/1, ST/W000628/1, ST/W000555/1, ST/W000547/1, ST/W00058X/1, ST/X508263/1, ST/V506862/1, ST/X508561/1, ST/V507040/1, ST/W507787/1, ST/R003181/1, ST/R003181/2,  ST/W507957/1, ST/X005984/1, ST/X006050/1; Portuguese Foundation for Science and Technology (FCT) under award numbers PTDC/FIS-PAR/2831/2020; the Institute for Basic Science, Korea (budget number IBS-R016-D1); the Swiss National Science Foundation (SNSF)  under award number 10001549. This research was supported by the Australian Government through the Australian Research Council Centre of Excellence for Dark Matter Particle Physics under award number CE200100008. We acknowledge additional support from the UK Science \& Technology Facilities Council (STFC) for PhD studentships and the STFC Boulby Underground Laboratory in the U.K., the GridPP~\cite{faulkner2005gridpp,britton2009gridpp} and IRIS Collaborations, in particular at Imperial College London and additional support by the University College London (UCL) Cosmoparticle Initiative, and the University of Zurich. We acknowledge additional support from the Center for the Fundamental Physics of the Universe, Brown University. K.T. Lesko acknowledges the support of Brasenose College and Oxford University. This research used resources of the National Energy Research Scientific Computing Center, a DOE Office of Science User Facility supported by the Office of Science of the U.S. Department of Energy under Contract No. DE-AC02-05CH11231. We gratefully acknowledge support from GitLab through its GitLab for Education Program. The University of Edinburgh is a charitable body, registered in Scotland, with the registration number SC005336. The assistance of SURF and its personnel in providing physical access and general logistical and technical support is acknowledged. We acknowledge the South Dakota Governor's office, the South Dakota Community Foundation, the South Dakota State University Foundation, and the University of South Dakota Foundation for use of xenon. We also acknowledge the University of Alabama for providing xenon. For the purpose of open access, the authors have applied a Creative Commons Attribution (CC BY) license to any Author Accepted Manuscript version arising from this submission. Finally, we respectfully acknowledge that we are on the traditional land of Indigenous American peoples and honor their rich cultural heritage and enduring contributions. Their deep connection to this land and their resilience and wisdom continue to inspire and enrich our community. We commit to learning from and supporting their effort as original stewards of this land and to preserve their cultures and rights for a more inclusive and sustainable future.


\appendix
\section{Appendix - Spin Dependent DM Results}

The treatment of spin-dependent models follows the procedure in Refs.~\cite{SR1WS,LZ:2024WIMP}. Refs.~\cite{PhysRevD.102.074018,Pirinen:2019gap,PhysRevLett.128.072502} describe the neutron-only and proton-only nuclear structure functions, with the structure functions from~\cite{PhysRevD.102.074018} treated as the nominal model. These structure functions are used as inputs to model DM scattering on \XeOneTwoNine~and \XeOneThreeOne.

\begin{figure*}[t]
	\centering
	{\includegraphics[width=0.48\textwidth]{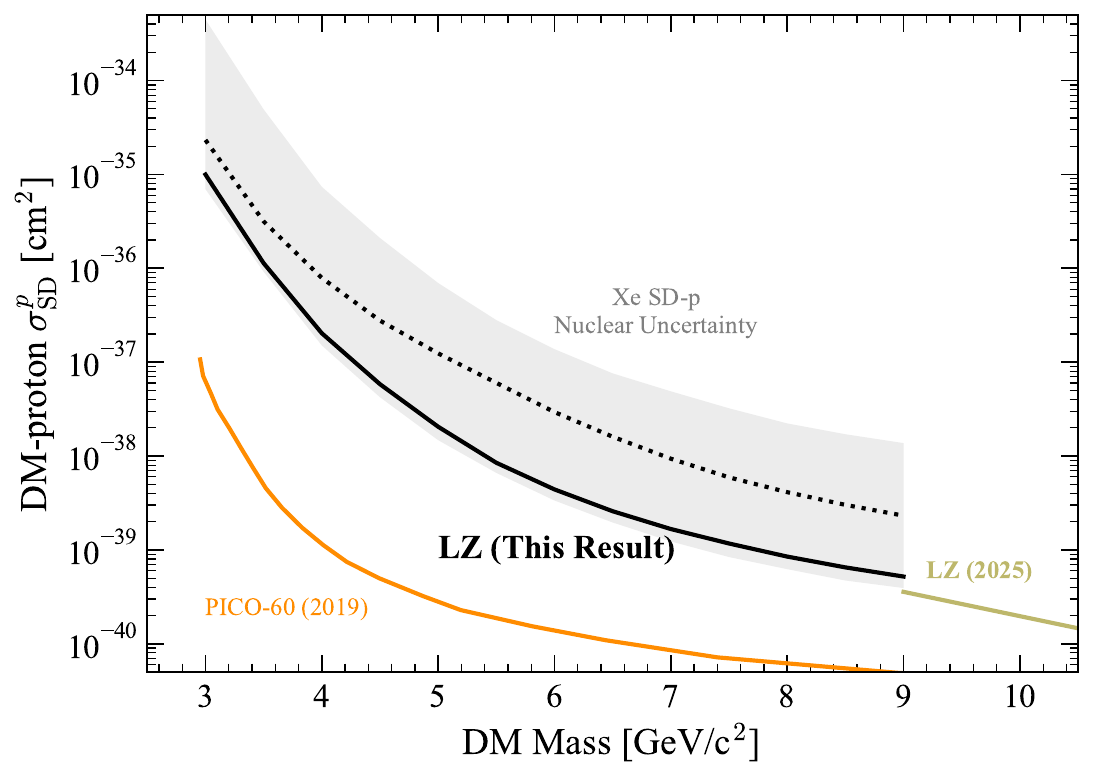}}\hfill
    {\includegraphics[width=0.48\textwidth]{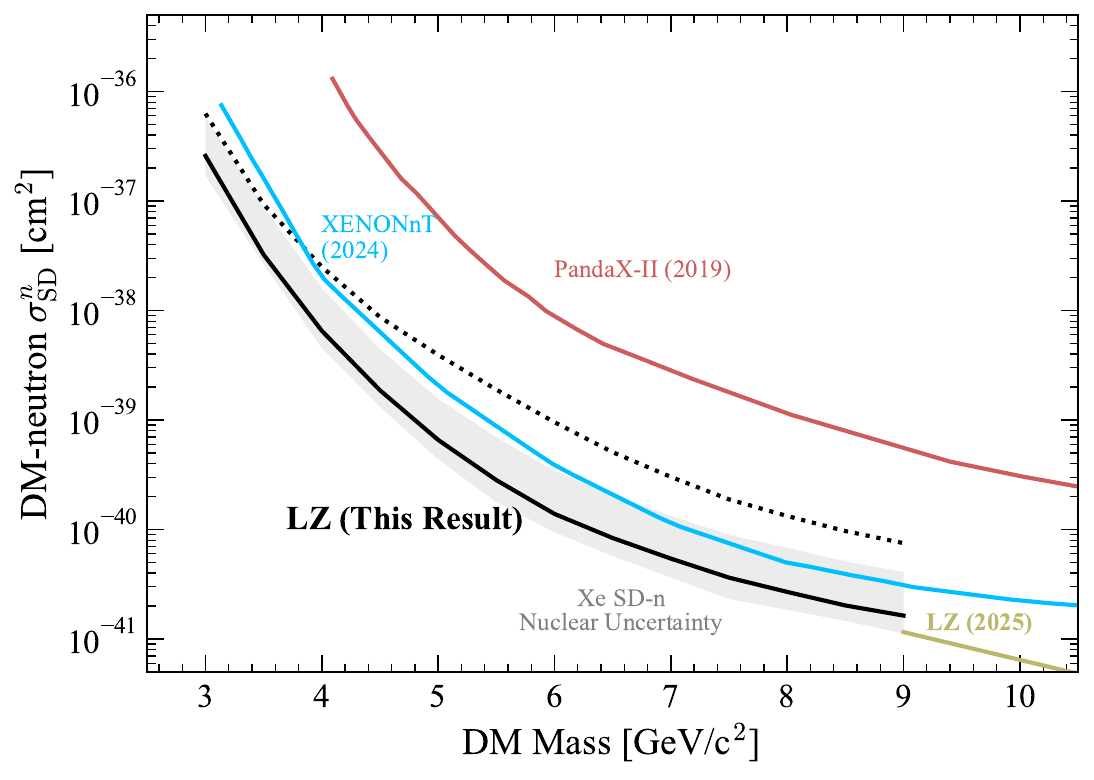}}
	\caption{Upper limits (90\% C.L.) from the 5.7 tonne-year analysis on the spin-dependent DM-proton (left) and DM-neutron (right) cross-section as a function of DM mass are shown with a solid black line. The observed limit uses the mean of the nuclear structure functions from~\cite{PhysRevD.102.074018}, with the gray band depicting the uncertainty from nuclear modeling in the models described in ~\cite{PhysRevD.102.074018,Pirinen:2019gap,PhysRevLett.128.072502}. The median 3$\sigma$ discovery potential using the post-fit model is shown as the dotted black line. Other experimental limits for spin-dependent DM-proton~\cite{amole2019dark,LZ:2024WIMP} and DM-neutron~\cite{LZ:2024WIMP,XENON:2023cxc} scattering are also shown.}
	\label{fig:SDLimits}
\end{figure*}

Upper limits (90\% confidence-level) on the spin-dependent cross-section as a function of DM mass are shown in Figure~\ref{fig:SDLimits} in both the DM-neutron and DM-proton coupling scenarios with the nuclear structure function uncertainty depicted as the gray band. The dotted black line depicts the 3$\sigma$ discovery potential. The strongest limit is set for a DM mass of 9~GeV/$c^2$ at a cross section of $\sigma_{\mathrm{SD}}^{n} =$ 1.6$\times10^{-41}$~cm$^{2}$ in the DM-neutron scenario, and at 9~GeV/$c^2$ at a cross section of $\sigma_{\mathrm{SD}}^p =$ 5.2$\times10^{-40}$~cm$^{2}$ in the DM-proton scenario.

\pagebreak

\bibliographystyle{apsrev4-2}
\bibliography{references}

@article{LZ-backgrounds,
    author = "Aalbers, J. and others",
    collaboration = "LZ",
    title = "{Background determination for the LUX-ZEPLIN dark matter experiment}",
    eprint = "2211.17120",
    archivePrefix = "arXiv",
    primaryClass = "hep-ex",
    doi = "10.1103/PhysRevD.108.012010",
    journal = "Phys. Rev. D",
    volume = "108",
    number = "1",
    pages = "012010",
    year = "2023"
}

@article{LZExperiment,
    author = "Akerib, D. S. and others",
    collaboration = "LZ",
    title = "{The LUX-ZEPLIN (LZ) Experiment}",
    eprint = "1910.09124",
    archivePrefix = "arXiv",
    primaryClass = "physics.ins-det",
    reportNumber = "FERMILAB-PUB-19-555-AE-E",
    doi = "10.1016/j.nima.2019.163047",
    journal = "Nucl. Instrum. Meth. A",
    volume = "953",
    pages = "163047",
    year = "2020"
}

@article{baxter2021recommended,
    author = "Baxter, D. and others",
    title = "{Recommended conventions for reporting results from direct dark matter searches}",
    eprint = "2105.00599",
    archivePrefix = "arXiv",
    primaryClass = "hep-ex",
    doi = "10.1140/epjc/s10052-021-09655-y",
    journal = "Eur. Phys. J. C",
    volume = "81",
    number = "10",
    pages = "907",
    year = "2021"
}

@article{cowan2011asymptotic,
    author = "Cowan, Glen and Cranmer, Kyle and Gross, Eilam and Vitells, Ofer",
    title = "{Asymptotic formulae for likelihood-based tests of new physics}",
    eprint = "1007.1727",
    archivePrefix = "arXiv",
    primaryClass = "physics.data-an",
    doi = "10.1140/epjc/s10052-011-1554-0",
    journal = "Eur. Phys. J. C",
    volume = "71",
    pages = "1554",
    year = "2011",
    note = "[Erratum: Eur.Phys.J.C 73, 2501 (2013)]"
}

@article{amole2019dark,
    author = "Amole, C. and others",
    collaboration = "PICO",
    title = "{Dark Matter Search Results from the Complete Exposure of the PICO-60 C$_3$F$_8$ Bubble Chamber}",
    eprint = "1902.04031",
    archivePrefix = "arXiv",
    primaryClass = "astro-ph.CO",
    reportNumber = "FERMILAB-PUB-19-073-AE-E",
    doi = "10.1103/PhysRevD.100.022001",
    journal = "Phys. Rev. D",
    volume = "100",
    number = "2",
    pages = "022001",
    year = "2019"
}

@misc{nest_2_4_0,
  author       = {Szydagis, M. and others},
  title        = {{Noble Element Simulation Technique, v2.4.0}},
  month        = August,
  year         = 2023,
  publisher    = {Zenodo},
  version      = {v2.4.0b},
  doi          = {https://doi.org/10.5281/zenodo.8215927},
  url          = {https://doi.org/10.5281/zenodo.8215927}
}

@misc{nest_2_4_5,
  author       = {Szydagis, M. and others},
  title        = {{Noble Element Simulation Technique, v2.4.5 beta}},
  month        = December,
  year         = 2025,
  publisher    = {Zenodo},
  version      = {v2.4.5beta},
  doi          = {https://doi.org/10.5281/zenodo.17851406},
  url          = {https://doi.org/10.5281/zenodo.17851406}
}

@article{akerib2021simulations,
    author = "Akerib, D. S. and others",
    collaboration = "LZ",
    title = "{Simulations of Events for the LUX-ZEPLIN (LZ) Dark Matter Experiment}",
    eprint = "2001.09363",
    archivePrefix = "arXiv",
    primaryClass = "physics.ins-det",
    reportNumber = "FERMILAB-PUB-20-401-AE",
    doi = "10.1016/j.astropartphys.2020.102480",
    journal = "Astropart. Phys.",
    volume = "125",
    pages = "102480",
    year = "2021"
}

@article{szydagis2022review,
    author = "Szydagis, M. and others",
    title = "{A review of NEST models for liquid xenon and an exhaustive comparison with other approaches}",
    eprint = "2211.10726",
    archivePrefix = "arXiv",
    primaryClass = "hep-ex",
    doi = "10.3389/fdest.2024.1480975",
    journal = "Front. Detect. Sci. Tech.",
    volume = "2",
    pages = "1480975",
    year = "2024"
}

@article{aharmin2013combined,
    author = "Aharmim, B. and others",
    collaboration = "SNO",
    title = "{Combined Analysis of all Three Phases of Solar Neutrino Data from the Sudbury Neutrino Observatory}",
    eprint = "1109.0763",
    archivePrefix = "arXiv",
    primaryClass = "nucl-ex",
    doi = "10.1103/PhysRevC.88.025501",
    journal = "Phys. Rev. C",
    volume = "88",
    pages = "025501",
    year = "2013"
}

@article{lztdr,
    author = "Mount, B. J. and others",
    title = "{LUX-ZEPLIN (LZ) Technical Design Report}",
    eprint = "1703.09144",
    archivePrefix = "arXiv",
    primaryClass = "physics.ins-det",
    reportNumber = "LBNL-1007256, FERMILAB-TM-2653-AE-E-PPD",
    month = "3",
    year = "2017",
    journal = ""
}

@article{faulkner2005gridpp,
  title={GridPP: development of the UK computing Grid for particle physics},
  author={Faulkner, PJW and others},
  journal={J. Phys. G},
  volume={32},
  number={1},
  pages={N1},
  year={2006},
  publisher={IOP Publishing},
  doi={https://doi.org/10.1088/0954-3899/32/1/N01}
}

@article{britton2009gridpp,
  title={GridPP: the UK grid for particle physics},
  author={Britton, David and others},
  journal={Philos. Trans.  R. Soc. A},
  volume={367},
  number={1897},
  pages={2447--2457},
  year={2009},
  publisher={The Royal Society London},
  doi={https://doi.org/10.1098/rsta.2009.0036}
}

@article{OHare:2021utq,
    author = "O'Hare, Ciaran A. J.",
    title = "{New Definition of the Neutrino Floor for Direct Dark Matter Searches}",
    eprint = "2109.03116",
    archivePrefix = "arXiv",
    primaryClass = "hep-ph",
    doi = "10.1103/PhysRevLett.127.251802",
    journal = "Phys. Rev. Lett.",
    volume = "127",
    number = "25",
    pages = "251802",
    year = "2021"
}

@article{akerib2022snowmass2021,
    author = "Akerib, D. S. and others",
    title = "{Snowmass2021 Cosmic Frontier Dark Matter Direct Detection to the Neutrino Fog}",
    booktitle = "{Snowmass 2021}",
    eprint = "2203.08084",
    archivePrefix = "arXiv",
    primaryClass = "hep-ex",
    reportNumber = "FERMILAB-CONF-22-180-V",
    month = "3",
    year = "2022",
    journal = ""
}

@article{Billard2022,
    author = "Billard, Julien and others",
    title = "{Direct detection of dark matter{\textemdash}APPEC committee report*}",
    eprint = "2104.07634",
    archivePrefix = "arXiv",
    primaryClass = "hep-ex",
    doi = "10.1088/1361-6633/ac5754",
    journal = "Rept. Prog. Phys.",
    volume = "85",
    number = "5",
    pages = "056201",
    year = "2022"
}

@article{Billard:2013qya,
    author = "Billard, J. and Strigari, L. and Figueroa-Feliciano, E.",
    title = "{Implication of neutrino backgrounds on the reach of next generation dark matter direct detection experiments}",
    eprint = "1307.5458",
    archivePrefix = "arXiv",
    primaryClass = "hep-ph",
    doi = "10.1103/PhysRevD.89.023524",
    journal = "Phys. Rev. D",
    volume = "89",
    number = "2",
    pages = "023524",
    year = "2014"
}

@article{Ruppin:2014bra,
    author = "Ruppin, F. and Billard, J. and Figueroa-Feliciano, E. and Strigari, L.",
    title = "{Complementarity of dark matter detectors in light of the neutrino background}",
    eprint = "1408.3581",
    archivePrefix = "arXiv",
    primaryClass = "hep-ph",
    doi = "10.1103/PhysRevD.90.083510",
    journal = "Phys. Rev. D",
    volume = "90",
    number = "8",
    pages = "083510",
    year = "2014"
}

@article{Bahcall:2004pz,
    author = "Bahcall, John N. and Serenelli, Aldo M. and Basu, Sarbani",
    title = "{New solar opacities, abundances, helioseismology, and neutrino fluxes}",
    eprint = "astro-ph/0412440",
    archivePrefix = "arXiv",
    doi = "10.1086/428929",
    journal = "Astrophys. J. Lett.",
    volume = "621",
    pages = "L85--L88",
    year = "2005"
}

@article{LUX_DD,
    author = "Akerib, D. S. and others",
    collaboration = "LUX",
    title = "{Low-energy (0.7-74 keV) nuclear recoil calibration of the LUX dark matter experiment using D-D neutron scattering kinematics}",
    eprint = "1608.05381",
    archivePrefix = "arXiv",
    primaryClass = "physics.ins-det",
    month = "8",
    year = "2016",
    journal = ""
}

@article{LUX:2022qxb,
    author = "Akerib, D. S. and others",
    collaboration = "LUX",
    title = "{Nuclear Recoil Calibration at Sub-keV Energies in LUX and Its Impact on Dark Matter Search Sensitivity}",
    eprint = "2210.05859",
    archivePrefix = "arXiv",
    primaryClass = "physics.ins-det",
    doi = "10.1103/PhysRevLett.134.061002",
    journal = "Phys. Rev. Lett.",
    volume = "134",
    number = "6",
    pages = "061002",
    year = "2025"
}

@article{PhysRevLett.128.072502,
    author = {Hu, B. S. and Padua-Arg{\"u}elles, J. and Leutheusser, S. and Miyagi, T. and Stroberg, S. R. and Holt, J. D.},
    title = "{Ab~Initio Structure Factors for Spin-Dependent Dark Matter Direct Detection}",
    eprint = "2109.00193",
    archivePrefix = "arXiv",
    primaryClass = "nucl-th",
    doi = "10.1103/PhysRevLett.128.072502",
    journal = "Phys. Rev. Lett.",
    volume = "128",
    number = "7",
    pages = "072502",
    year = "2022"
}

@article{PhysRevD.102.074018,
    author = "Hoferichter, Martin and Men{\'e}ndez, Javier and Schwenk, Achim",
    title = "{Coherent elastic neutrino-nucleus scattering: EFT analysis and nuclear responses}",
    eprint = "2007.08529",
    archivePrefix = "arXiv",
    primaryClass = "hep-ph",
    reportNumber = "INT-PUB-20-026",
    doi = "10.1103/PhysRevD.102.074018",
    journal = "Phys. Rev. D",
    volume = "102",
    number = "7",
    pages = "074018",
    year = "2020"
}

@ARTICLE{Pirinen:2019gap,
       author = {{Pirinen}, P. and {Kotila}, J. and {Suhonen}, J.},
        title = "{Spin-dependent WIMP-nucleus scattering off $^{125}$Te, $^{129}$Xe, and $^{131}$Xe in the microscopic interacting boson-fermion model}",
      journal = {\nphysa},
         year = 2019,
        month = dec,
       volume = {992},
          eid = {121624},
        pages = {121624},
          doi = {10.1016/j.nuclphysa.2019.121624},
       adsurl = {https://ui.adsabs.harvard.edu/abs/2019NuPhA.99221624P}
}

@article{SR1WS,
    author = "Aalbers, J. and others",
    collaboration = "LZ Collaboration",
    title = "{First Dark Matter Search Results from the LUX-ZEPLIN (LZ) Experiment}",
    eprint = "2207.03764",
    archivePrefix = "arXiv",
    primaryClass = "hep-ex",
    doi = "10.1103/PhysRevLett.131.041002",
    journal = "Phys. Rev. Lett.",
    volume = "131",
    number = "4",
    pages = "041002",
    year = "2023"
}

@article{XENON:2023cxc,
    author = "Aprile, E. and others",
    collaboration = "XENON Collaboration",
    title = "{First Dark Matter Search with Nuclear Recoils from the XENONnT Experiment}",
    eprint = "2303.14729",
    archivePrefix = "arXiv",
    primaryClass = "hep-ex",
    doi = "10.1103/PhysRevLett.131.041003",
    journal = "Phys. Rev. Lett.",
    volume = "131",
    number = "4",
    pages = "041003",
    year = "2023"
}

@article{LZ_calibrations,
    author = "Aalbers, J. and others",
    collaboration = "LZ",
    title = "{The design, implementation, and performance of the LZ calibration systems}",
    eprint = "2406.12874",
    archivePrefix = "arXiv",
    primaryClass = "physics.ins-det",
    doi = "10.1088/1748-0221/19/08/P08027",
    journal = "JINST",
    volume = "19",
    number = "08",
    pages = "P08027",
    year = "2024"
}

@article{LZ_DAQ,
    title = "{The Data Acquisition System of the LZ Dark Matter Detector: FADR}",
    journal = {Nucl. Instrum. Methods Phys. Res. Sect. A},
    volume = {1068},
    pages = {169712},
    year = {2024},
    issn = {0168-9002},
    doi = {https://doi.org/10.1016/j.nima.2024.169712},
    url = {https://www.sciencedirect.com/science/article/pii/S0168900224006387},
    eprint = "2405.14732",
    journal={arXiv preprint},
    author = "Aalbers, J. and others",
    collaboration = "LZ Collaboration",
}

@article{LZ:2024WIMP,
    author = "Aalbers, J. and others",
    collaboration = "LZ Collaboration",
    title = "{Dark Matter Search Results from 4.2{\,}{\,}Tonne-Years of Exposure of the LUX-ZEPLIN (LZ) Experiment}",
    eprint = "2410.17036",
    archivePrefix = "arXiv",
    primaryClass = "hep-ex",
    reportNumber = "FERMILAB-PUB-24-0796-V",
    doi = "10.1103/4dyc-z8zf",
    journal = "Phys. Rev. Lett.",
    volume = "135",
    number = "1",
    pages = "011802",
    year = "2025"
}

@misc{hansen2019pycma,
  author       = {Nikolaus Hansen and Youhei Akimoto and Petr Baudis},
  title        = {{CMA-ES/pycma} on {G}ithub},
  howpublished = {Zenodo, DOI:10.5281/zenodo.2559634},
  month        = feb,
  year         = 2019,
  doi          = {10.5281/zenodo.2559634},
  url          = {https://doi.org/10.5281/zenodo.2559634},
}

@article{Panda:8B,
    author = "Bo, Zihao and others",
    collaboration = "PandaX",
    title = "{First Indication of Solar B8 Neutrinos through Coherent Elastic Neutrino-Nucleus Scattering in PandaX-4T}",
    eprint = "2407.10892",
    archivePrefix = "arXiv",
    primaryClass = "hep-ex",
    doi = "10.1103/PhysRevLett.133.191001",
    journal = "Phys. Rev. Lett.",
    volume = "133",
    number = "19",
    pages = "191001",
    year = "2024"
}

@article{XenonNT:8B,
    author = "Aprile, Elena and others",
    collaboration = "XENON",
    title = "{First Indication of Solar B8 Neutrinos via Coherent Elastic Neutrino-Nucleus Scattering with XENONnT}",
    eprint = "2408.02877",
    archivePrefix = "arXiv",
    primaryClass = "nucl-ex",
    doi = "10.1103/PhysRevLett.133.191002",
    journal = "Phys. Rev. Lett.",
    volume = "133",
    number = "19",
    pages = "191002",
    year = "2024"
}

@article{XenonNT:LDM-NeutrinoFog,
    author = "Aprile, E. and others",
    collaboration = "XENON",
    title = "{First Search for Light Dark Matter in the Neutrino Fog with XENONnT}",
    eprint = "2409.17868",
    archivePrefix = "arXiv",
    primaryClass = "hep-ex",
    doi = "10.1103/PhysRevLett.134.111802",
    journal = "Phys. Rev. Lett.",
    volume = "134",
    number = "11",
    pages = "111802",
    year = "2025"
}

@article{Maity:2024aji,
    author = "Maity, Tarak Nath and Boehm, Celine",
    title = "{First constraint on the weak mixing angle using direct detection experiments}",
    eprint = "2409.04385",
    archivePrefix = "arXiv",
    primaryClass = "hep-ph",
    doi = "10.1103/1lgl-bx7v",
    journal = "Phys. Rev. D",
    volume = "112",
    number = "5",
    pages = "053001",
    year = "2025"
}

@article{Chadwick20112887short,
    title = "{ENDF/B-VII.1} Nuclear Data for Science and Technology: Cross Sections, Covariances, Fission Product Yields and Decay Data",
    journal = "Nuclear Data Sheets",
    volume = "112",
    number = "12",
    pages = "2887 - 2996",
    year = "2011",
    note = " Special Issue on ENDF/B-VII.1 Library ",
    issn = "0090-3752",
    doi = "10.1016/j.nds.2011.11.002",
    url = "https://www.sciencedirect.com/science/article/pii/S009037521100113X",
    author = "M.B. Chadwick and M. Herman and P. Oblo{\v z}insk{\' y} and others" 
}

@ARTICLE{G4NDL,
  author={Mendoza, Emilio and Cano-Ott, Daniel and Koi, Tatsumi and Guerrero, Carlos},
  journal={IEEE Transactions on Nuclear Science}, 
  title={New Standard Evaluated Neutron Cross Section Libraries for the GEANT4 Code and First Verification}, 
  year={2014},
  volume={61},
  number={4},
  pages={2357-2364},
  doi={10.1109/TNS.2014.2335538}}

@article{BAKER1984437,
title = {Clarification of the use of CHI-square and likelihood functions in fits to histograms},
journal = {Nuclear Instruments and Methods in Physics Research},
volume = {221},
number = {2},
pages = {437-442},
year = {1984},
issn = {0167-5087},
doi = {https://doi.org/10.1016/0167-5087(84)90016-4},
url = {https://www.sciencedirect.com/science/article/pii/0167508784900164},
author = {Steve Baker and Robert D. Cousins}
}

@phdthesis{rhynethesis,
    title        = {Development of Tagged Low Energy D- and H-Based Scintillator Reflector Neutron Sources and Characterizations of a D-D Neutron Generator for Calibrations of the LZ Detector},
    author       = {Casey Rhyne},
    year         = 2024,
    url = {https://repository.library.brown.edu/studio/item/bdr:eqjnr8x6/PDF/},
    school       = {Brown University},
    type         = {{PhD Thesis}}
}

@ARTICLE{emcee,
       author = {{Foreman-Mackey}, Daniel and {Hogg}, David W. and {Lang}, Dustin and {Goodman}, Jonathan},
        title = "{emcee: The MCMC Hammer}",
      journal = {\pasp},
         year = 2013,
        month = mar,
       volume = {125},
       number = {925},
        pages = {306},
          doi = {10.1086/670067},
archivePrefix = {arXiv},
       eprint = {1202.3665},
 primaryClass = {astro-ph.IM},
       adsurl = {https://ui.adsabs.harvard.edu/abs/2013PASP..125..306F}
}

@article{Akerib:2021pfd,
    author = "Akerib, D. S. and others",
    title = "{Enhancing the sensitivity of the LUX-ZEPLIN (LZ) dark matter experiment to low energy signals}",
    eprint = "2101.08753",
    archivePrefix = "arXiv",
    primaryClass = "astro-ph.IM",
    month = "1",
    year = "2021",
    journal = ""
}

@article{Akerib:2025rzt,
    author = "Akerib, D. S. and others",
    title = "{Study of few-electron backgrounds in the LUX-ZEPLIN detector}",
    eprint = "2510.06500",
    archivePrefix = "arXiv",
    primaryClass = "physics.ins-det",
    month = "10",
    year = "2025",
    journal = ""
}

@article{LZ:2025vdo,
    author = "Aalbers, J. and others",
    collaboration = "LZ Collaboration",
    title = "{Low-energy nuclear recoil calibration of the LUX-ZEPLIN experiment with a photoneutron source}",
    eprint = "2509.16281",
    archivePrefix = "arXiv",
    primaryClass = "physics.ins-det",
    month = "9",
    year = "2025",
    journal = ""
}

@article{XENON:2024kbh,
    author = "Aprile, E. and others",
    collaboration = "XENON Collaboration",
    title = "{Low-Energy Nuclear Recoil Calibration of XENONnT with a $^{88}$YBe Photoneutron Source}",
    eprint = "2412.10451",
    archivePrefix = "arXiv",
    primaryClass = "physics.ins-det",
    month = "12",
    year = "2024",
    journal = ""
}

@article{Lenardo:2019fcn,
    author = "Lenardo, Brian and others",
    title = "{Measurement of the ionization yield from nuclear recoils in liquid xenon between 0.3 - 6 keV with single-ionization-electron sensitivity}",
    eprint = "1908.00518",
    archivePrefix = "arXiv",
    primaryClass = "physics.ins-det",
    month = "8",
    year = "2019",
    journal = ""
}

@article{nerix:2018,
    author = "Aprile, E. and Anthony, M. and Lin, Q. and Greene, Z. and De Perio, P. and Gao, F. and Howlett, J. and Plante, G. and Zhang, Y. and Zhu, T.",
    title = "{Simultaneous measurement of the light and charge response of liquid xenon to low-energy nuclear recoils at multiple electric fields}",
    eprint = "1809.02072",
    archivePrefix = "arXiv",
    primaryClass = "physics.ins-det",
    doi = "10.1103/PhysRevD.98.112003",
    journal = "Phys. Rev. D",
    volume = "98",
    number = "11",
    pages = "112003",
    year = "2018"
}

@article{PandaX:2022aac,
    author = "Ma, Wenbo and others",
    collaboration = "PandaX Collaboration",
    title = "{Search for Solar B8 Neutrinos in the PandaX-4T Experiment Using Neutrino-Nucleus Coherent Scattering}",
    eprint = "2207.04883",
    archivePrefix = "arXiv",
    primaryClass = "hep-ex",
    doi = "10.1103/PhysRevLett.130.021802",
    journal = "Phys. Rev. Lett.",
    volume = "130",
    number = "2",
    pages = "021802",
    year = "2023"
}

@article{PandaX2025,
    author = "Zhang, Minzhen and others",
    collaboration = "PandaX",
    title = "{Search for Light Dark Matter with 259 Days of Data in PandaX-4T}",
    eprint = "2507.11930",
    archivePrefix = "arXiv",
    primaryClass = "hep-ex",
    doi = "10.1103/rtnh-jn8s",
    journal = "Phys. Rev. Lett.",
    volume = "135",
    number = "21",
    pages = "211001",
    year = "2025"
}

@article{DarkSide-50:2022qzh,
    author = "Agnes, P. and others",
    collaboration = "DarkSide-50 Collaboration",
    title = "{Search for low-mass dark matter WIMPs with 12~ton-day exposure of DarkSide-50}",
    eprint = "2207.11966",
    archivePrefix = "arXiv",
    primaryClass = "hep-ex",
    reportNumber = "FERMILAB-PUB-22-589-ND-PPD-SCD",
    doi = "10.1103/PhysRevD.107.063001",
    journal = "Phys. Rev. D",
    volume = "107",
    number = "6",
    pages = "063001",
    year = "2023"
}

@article{Vinyoles:2016djt,
    author = {Vinyoles, N{\'u}ria and others},
    title = "{A new Generation of Standard Solar Models}",
    eprint = "1611.09867",
    archivePrefix = "arXiv",
    primaryClass = "astro-ph.SR",
    doi = "10.3847/1538-4357/835/2/202",
    journal = "Astrophys. J.",
    volume = "835",
    number = "2",
    pages = "202",
    year = "2017"
}

@article{ParticleDataGroup:2024cfk,
    author = "Navas, S. and others",
    collaboration = "Particle Data Group",
    title = "{Review of particle physics}",
    doi = "10.1103/PhysRevD.110.030001",
    journal = "Phys. Rev. D",
    volume = "110",
    number = "3",
    pages = "030001",
    year = "2024"
}

@article{Wood_1997_APV,
author = {C. S. Wood  and S. C. Bennett  and D. Cho  and B. P. Masterson  and J. L. Roberts  and C. E. Tanner  and C. E. Wieman },
title = {Measurement of Parity Nonconservation and an Anapole Moment in Cesium},
journal = {Science},
volume = {275},
number = {5307},
pages = {1759-1763},
year = {1997}
}

@article{Anthony_2005_SLAC_E158,
    author = "Anthony, P. L. and others",
    collaboration = "SLAC E158",
    title = "{Precision measurement of the weak mixing angle in Moller scattering}",
    eprint = "hep-ex/0504049",
    archivePrefix = "arXiv",
    reportNumber = "SLAC-PUB-11149",
    doi = "10.1103/PhysRevLett.95.081601",
    journal = "Phys. Rev. Lett.",
    volume = "95",
    pages = "081601",
    year = "2005"
}

@article{Androic_2018_Qweak,
    author = "Androi{\'c}, D. and others",
    collaboration = "Qweak",
    title = "{Precision measurement of the weak charge of the proton}",
    eprint = "1905.08283",
    archivePrefix = "arXiv",
    primaryClass = "nucl-ex",
    doi = "10.1038/s41586-018-0096-0",
    journal = "Nature",
    volume = "557",
    number = "7704",
    pages = "207--211",
    year = "2018"
}

@article{DeRomeri_2023_COHERENT,
    author = "De Romeri, V. and Miranda, O. G. and Papoulias, D. K. and Sanchez Garcia, G. and T{\'o}rtola, M. and Valle, J. W. F.",
    title = "{Physics implications of a combined analysis of COHERENT CsI and LAr data}",
    eprint = "2211.11905",
    archivePrefix = "arXiv",
    primaryClass = "hep-ph",
    doi = "10.1007/JHEP04(2023)035",
    journal = "JHEP",
    volume = "04",
    pages = "035",
    year = "2023"
}

@article{Sierra_2022_DresdenII,
    author = "Aristizabal Sierra, D. and De Romeri, V. and Papoulias, D. K.",
    title = "{Consequences of the Dresden-II reactor data for the weak mixing angle and new physics}",
    eprint = "2203.02414",
    archivePrefix = "arXiv",
    primaryClass = "hep-ph",
    doi = "10.1007/JHEP09(2022)076",
    journal = "JHEP",
    volume = "09",
    pages = "076",
    year = "2022"
}

@article{DeRomeri_2025_CONUS,
    author = "De Romeri, Valentina and Papoulias, Dimitrios K. and Sanchez Garcia, Gonzalo",
    title = "{Implications of the first CONUS+ measurement of coherent elastic neutrino-nucleus scattering}",
    eprint = "2501.17843",
    archivePrefix = "arXiv",
    primaryClass = "hep-ph",
    doi = "10.1103/PhysRevD.111.075025",
    journal = "Phys. Rev. D",
    volume = "111",
    number = "7",
    pages = "075025",
    year = "2025"
}

@article{Prescott_1979_eDIS_Old,
title = {Further measurements of parity non-conservation in inelastic electron scattering},
journal = {Physics Letters B},
volume = {84},
number = {4},
pages = {524-528},
year = {1979},
author = {Prescott, C.Y. and others}
}

@phdthesis{huang,
    author = "Huang, Dongqing",
    title = "{Ultra-Low Energy Calibration of the LUX and LZ Dark Matter Detectors}",
    doi = "10.26300/zvs6-fx07",
    school = "Brown U.",
    year = "2020"
}

@article{opencv,
    author = {Bradski, G.},
    journal = {Dr. Dobb's Journal of Software Tools},
    title = {{The OpenCV Library}},
    year = {2000}
}

@article{Pershing_2022,
    author = "Pershing, Teal and others",
    title = "{Calibrating the scintillation and ionization responses of xenon recoils for high-energy dark matter searches}",
    eprint = "2207.08326",
    archivePrefix = "arXiv",
    primaryClass = "physics.ins-det",
    doi = "10.1103/PhysRevD.106.052013",
    journal = "Phys. Rev. D",
    volume = "106",
    number = "5",
    pages = "052013",
    year = "2022"
}

@article{Kaplan:2009ag,
    author = "Kaplan, David E. and Luty, Markus A. and Zurek, Kathryn M.",
    title = "{Asymmetric Dark Matter}",
    eprint = "0901.4117",
    archivePrefix = "arXiv",
    primaryClass = "hep-ph",
    reportNumber = "FERMILAB-PUB-09-345-A-T",
    doi = "10.1103/PhysRevD.79.115016",
    journal = "Phys. Rev. D",
    volume = "79",
    pages = "115016",
    year = "2009"
}

@article{Cohen:2010kn,
    author = "Cohen, Timothy and Phalen, Daniel J. and Pierce, Aaron and Zurek, Kathryn M.",
    title = "{Asymmetric Dark Matter from a GeV Hidden Sector}",
    eprint = "1005.1655",
    archivePrefix = "arXiv",
    primaryClass = "hep-ph",
    reportNumber = "MCTP-10-18",
    doi = "10.1103/PhysRevD.82.056001",
    journal = "Phys. Rev. D",
    volume = "82",
    pages = "056001",
    year = "2010"
}

@article{Alpizar:2025wor,
    author = "Alp{\'\i}zar-Venegas, M. and Flores, L. J. and Peinado, Eduardo and V{\'a}zquez-J{\'a}uregui, E.",
    title = "{Exploring the standard model and beyond from the evidence of CE{\ensuremath{\nu}}NS with reactor antineutrinos in CONUS+}",
    eprint = "2501.10355",
    archivePrefix = "arXiv",
    primaryClass = "hep-ph",
    doi = "10.1103/PhysRevD.111.053001",
    journal = "Phys. Rev. D",
    volume = "111",
    number = "5",
    pages = "053001",
    year = "2025"
}

@ARTICLE{Freedman_1977,
       author = {{Freedman}, D.~Z. and {Schramm}, D.~N. and {Tubbs}, D.~L.},
        title = "{The Weak Neutral Current and its Effects in Stellar Collapse}",
      journal = {Annual Review of Nuclear and Particle Science},
         year = 1977,
        month = jan,
       volume = {27},
        pages = {167-207},
          doi = {10.1146/annurev.ns.27.120177.001123},
       adsurl = {https://ui.adsabs.harvard.edu/abs/1977ARNPS..27..167F}
}

@article{Freedman_1974,
  title = {Coherent effects of a weak neutral current},
  author = {Freedman, Daniel Z.},
  journal = {Phys. Rev. D},
  volume = {9},
  issue = {5},
  pages = {1389--1392},
  numpages = {0},
  year = {1974},
  month = {Mar},
  publisher = {American Physical Society},
  doi = {10.1103/PhysRevD.9.1389},
  url = {https://link.aps.org/doi/10.1103/PhysRevD.9.1389}
}

@article{Bahcall_1996,
  title = {Standard neutrino spectrum from $^{8}\mathrm{B}$ decay},
  author = {Bahcall, John N. and Lisi, E. and Alburger, D. E. and De Braeckeleer, L. and Freedman, S. J. and Napolitano, J.},
  journal = {Phys. Rev. C},
  volume = {54},
  issue = {1},
  pages = {411--422},
  numpages = {0},
  year = {1996},
  month = {Jul},
  publisher = {American Physical Society},
  doi = {10.1103/PhysRevC.54.411},
  url = {https://link.aps.org/doi/10.1103/PhysRevC.54.411}
}

@article{Longfellow_2023,
  title = {Determination of the $^{8}\mathrm{B}$ neutrino energy spectrum using trapped ions},
  author = {Longfellow, B. and others},
  journal = {Phys. Rev. C},
  volume = {107},
  issue = {3},
  pages = {L032801},
  numpages = {6},
  year = {2023},
  month = {Mar},
  publisher = {American Physical Society},
  doi = {10.1103/PhysRevC.107.L032801},
  url = {https://link.aps.org/doi/10.1103/PhysRevC.107.L032801}
}

@article{Tomalak_2021,
   title={Flavor-dependent radiative corrections in coherent elastic neutrino-nucleus scattering},
   volume={2021},
   ISSN={1029-8479},
   url={http://dx.doi.org/10.1007/JHEP02(2021)097},
   DOI={10.1007/jhep02(2021)097},
   number={2},
   journal={Journal of High Energy Physics},
   publisher={Springer Science and Business Media LLC},
   author={Tomalak, Oleksandr and Machado, Pedro and Pandey, Vishvas and Plestid, Ryan},
   year={2021},
   month=Feb }

@article{Erler_2013,
   title={The weak neutral current},
   volume={71},
   ISSN={0146-6410},
   url={http://dx.doi.org/10.1016/j.ppnp.2013.03.004},
   DOI={10.1016/j.ppnp.2013.03.004},
   journal={Progress in Particle and Nuclear Physics},
   publisher={Elsevier BV},
   author={Erler, Jens and Su, Shufang},
   year={2013},
   month=July, pages={119–149} }

@article{Marciano_1980,
  title = {Radiative corrections to neutrino-induced neutral-current phenomena in the $\mathrm{SU}{(2)}_{L}\ifmmode\times\else\texttimes\fi{}\mathrm{U}(1)$ theory},
  author = {Marciano, W. J. and Sirlin, A.},
  journal = {Phys. Rev. D},
  volume = {22},
  issue = {11},
  pages = {2695--2717},
  numpages = {0},
  year = {1980},
  month = {Dec},
  publisher = {American Physical Society},
  doi = {10.1103/PhysRevD.22.2695},
  url = {https://link.aps.org/doi/10.1103/PhysRevD.22.2695}
}

@article{Klein_and_Roodman_2005,
   author = "Klein, Joshua R and Roodman, Aaron",
   title = "BLIND ANALYSIS IN NUCLEAR AND PARTICLE PHYSICS", 
   journal= "Annual Review of Nuclear and Particle Science",
   year = "2005",
   volume = "55",
   number = "Volume 55, 2005",
   pages = "141-163",
   doi = "https://doi.org/10.1146/annurev.nucl.55.090704.151521",
   url = "https://www.annualreviews.org/content/journals/10.1146/annurev.nucl.55.090704.151521",
   publisher = "Annual Reviews",
   issn = "1545-4134",
   type = "Journal Article"
  }

@article{Sanchez_2003,
  title = {Measurement of the $L/E$ distributions of atmospheric $\ensuremath{\nu}$ in Soudan 2 and their interpretation as neutrino oscillations},
  author = {Sanchez, M. and others},
  journal = {Phys. Rev. D},
  volume = {68},
  issue = {11},
  pages = {113004},
  numpages = {14},
  year = {2003},
  month = {Dec},
  publisher = {American Physical Society},
  doi = {10.1103/PhysRevD.68.113004},
  url = {https://link.aps.org/doi/10.1103/PhysRevD.68.113004}
}

@article{Savage_1986,
title = {A search for fractional charges in native mercury},
journal = {Physics Letters B},
volume = {167},
number = {4},
pages = {481-484},
year = {1986},
issn = {0370-2693},
doi = {https://doi.org/10.1016/0370-2693(86)91305-5},
url = {https://www.sciencedirect.com/science/article/pii/0370269386913055},
author = {Maureen L. Savage and others}
}

@phdthesis{rjamesthesis,
    title        = {Signals, backgrounds and statistical inference for dark matter direct detection experiments},
    author       = {Robert James},
    year         = 2024,
    url = {https://discovery.ucl.ac.uk/id/eprint/10190030/},
    school       = {University College London},
    type         = {{PhD Thesis}}
}

\clearpage
\pagebreak

\widetext
\begin{center}
\textbf{\large Supplemental Materials}
\end{center}
\setcounter{equation}{0}
\setcounter{figure}{0}
\setcounter{table}{0}
\setcounter{page}{1}
\makeatletter
\renewcommand{\theequation}{S\arabic{equation}}
\renewcommand{\thefigure}{S\arabic{figure}}
\renewcommand{\thetable}{S\arabic{table}}

\section{Details of the Accidental Coincidence Model and Validation}\label{sup:ac} 

We construct the accidental coincidence model with two major ingredients: a high statistics synthetic dataset generated from real data and a sideband of single scatter events with unphysically long drift times to constrain the overall rate. To produce the synthetic dataset, we draw S1 and S2 pulses from non-scatter events which pass a minimal set of data quality criteria and loose pulse selections. We then combine the pulses and their surrounding electron and photon environments together at the waveform level to form accidental coincidence events. This procedure produces synthetic events with both physical drift times ($\lesssim$~1~ms) and unphysical drift times ($\gtrsim$~1~ms). 

We normalize the model by scaling the rate of synthetic unphysical drift time (UDT) events to a sideband dataset consisting of UDT events. This dataset spans a broad range of pulse areas, inclusive of the ROI, with S1s between 2 and 15~phd and S2s up to $10^4$~phd. We impose a limited set of data selections to maintain sufficient statistics; namely, the S2 ROI, fiducial volume, and several S1 and S2 pulse-based selections are not applied.

After normalization, we apply the remaining data selection criteria to the synthetic physical drift time events to produce the final accidental coincidence prediction in \{\sonec,~\stwoc\}. We use the synthetic dataset to derive the predicted ratio of UDT to physical drift time (PDT) events at normalization since only PDT events pass the fiducial volume selection. We validate this ratio with an ancillary dataset of events failing S1 and S2 pulse-based selections. When applying the remaining selections, we also confirm that the fraction of UDT events which pass each selection in sequence is consistent between the model and sideband dataset. All comparisons agree to within $1.5\sigma$. 

\begin{figure}[!h]
	\centering
	\includegraphics[width=0.6 \columnwidth]{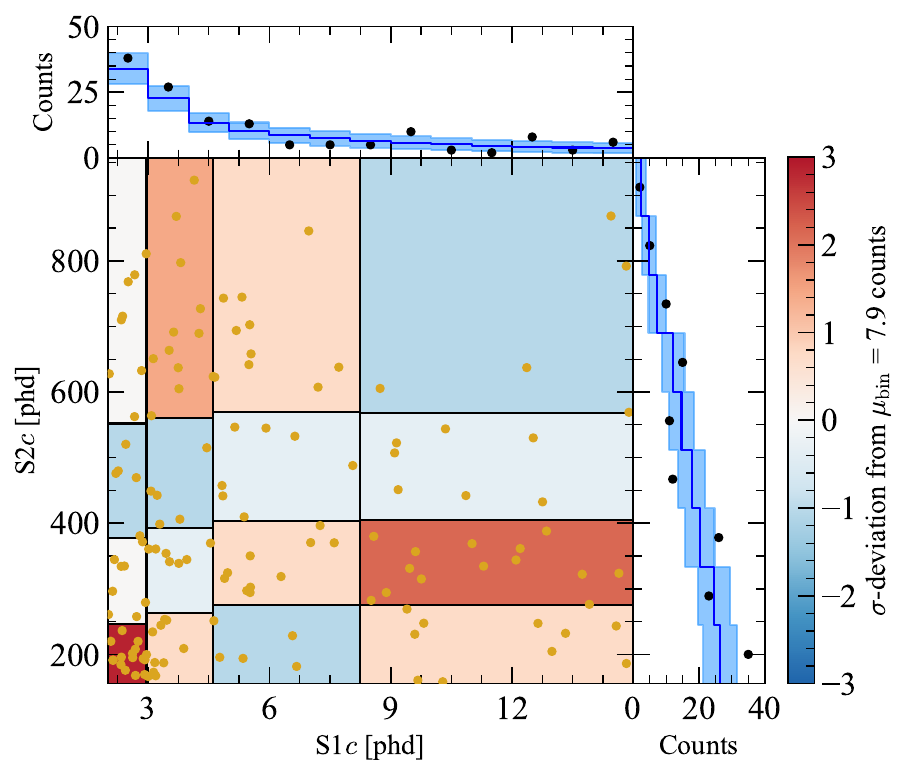}
	\caption{Data-model agreement for a pure accidental background validation dataset composed of events with unphysical drift time ($\gtrsim$~1~ms), shown in \{\sonec,~\stwoc\} and 1D projections. In this comparison, a subset of selection criteria applied to the final dataset is removed for increased statistics. The underlying color plot shows good data-model agreement in units of $\sigma$. These bins are defined such that in each the model predicts 7.9 events. The yellow and black dots depict the data, while the blue line and band depict the accidental model prediction and associated Poisson uncertainty.}
	\label{fig:Acc}
\end{figure}

Figure~\ref{fig:Acc} shows the data-model agreement for UDT events restricted to the ROI prior to application of the fiducial volume and the aforementioned S1 and S2 pulse-based selections. The model shows excellent agreement with the data, with a p-value from binned goodness-of-fit tests of 0.28 in \{\sonec,~\stwoc\} and 0.38, 0.48 in the S1$c$, S2$c$ projections, respectively.

We test the robustness of this modeling algorithm with ancillary datasets which probe a range of environmental conditions and assumptions. To test the modeling in elevated electron and photon rate environments, we loosen the hold-off after large energy depositions described in the main text which results in the inclusion of events with approximately 3 and 6 times the mean electron and photon rates in the main science dataset, respectively. We then compare the model to UDT events in this elevated electron and photon rate environment. We also perform the modeling procedure on events during AmLi calibration periods with elevated S1 and S2 pulse rates and check the agreement with UDT events. To test the modeling of PDT events, we compare the data-model agreement for physical drift time events that fail several S1 and S2 pulse cuts. More details on the model construction and validations will be described in a dedicated modeling paper.

\section{Details of the Nuclear Recoil Detector Response Model and Uncertainty Treatment}
\label{sup:NRmodel}

As described in the main text, the LZ nuclear recoil response model is based on \textsc{nest}~v2.4.5~beta~\cite{nest_2_4_5} which is tuned to 2.45~MeV DD-neutron-induced xenon nuclear recoils from threshold (1~keV) to the 74~keV endpoint. We employ a Markov Chain Monte Carlo (MCMC) procedure using the \textsc{emcee} package~\cite{emcee}. Table~\ref{tab:nest_params} shows the default values, priors assumed, and resulting best-fit values for the parameters governing the NR yields, NR fluctuations, and model-specific nuisance parameters. \textsc{nest} parameters excluded from the table, such as those describing field dependence (e.g. $\delta$), are held fixed at their default values due to a lack of sensitivity.

The yield parameters $\alpha$, $\beta$, $\gamma$, and $\epsilon$, as well as the fluctuation parameters $F_\text{i}$ and the NR recombination variance coefficient, $\lambda$, are given wide flat priors that encompass the \textsc{nest} default values. $F_\text{ex}$ is modified in the model to have a linear dependence on the number of excitons $N_\text{ex}$, with the functional form $F_\text{ex}=aN_\text{ex}$+$b$, with coefficients $a$ and $b$ floated in the fit. The modification of $F_\text{ex}$ from the \textsc{nest} default is found to be necessary for good data-model agreement. The parameters controlling the low energy roll-off behavior, $\zeta$, $\eta$, $\theta$, $\iota$, have Gaussian prior distributions in the MCMC, with the $\mu$ and $\sigma$ values set by the \textsc{nest} default mean values and uncertainties, respectively. This choice is motivated by the minimum energy at which this calibration is sensitive ($\sim2.5$~keV), determined by incrementally increasing the minimum recoil energy included in the model until the fit $\Delta\chi^2$ exceeds 1. These \textsc{nest} model priors are driven by yields measurements performed down to sub-keV NR energies~\cite{LUX_DD,LUX:2022qxb,Lenardo:2019fcn}. 

We further consider nuisance parameters specific to the physics of the neutron interactions. First, the 2.45~MeV DD neutrons may have their energy degraded in two independent processes due to a) interactions within the DD generator before exiting and b) interactions in intervening detector material before entering the active xenon volume. The DD neutron kinetic energy spectrum used in this analysis is based on dedicated simulations and measurements performed at Brown University prior to deploying the DD generator underground at SURF~\cite{rhynethesis}. These effects manifest in the fraction of $<$2~MeV neutrons considered in the energy spectrum entering the TPC. Second, we consider uncertainties in the xenon differential elastic-scattering cross-sections by treating JENDL-4.0u~\cite{G4NDL} and ENDF/B-VII.1~\cite{Chadwick20112887short} as limiting cases for the single-scatter NR spectrum. Both of these parameters primarily affect the observed recoil spectra at intermediate recoil energies (approximately from 25--50~keV), and are found to be largely uncorrelated with the mean yields and fluctuation parameters.

We study two background populations: accidental coincidences like those described in the text and unresolved multiple scatters - multiple scatter events where the S2s cannot be resolved in time and appear as single scatters. The accidental population is estimated using both UDT events and events where the S1 does not coincide with the DD generator pulse. It is found to have a small predicted rate and limited impact on the best fit yields and fluctuations; thus it is fixed in the fits. The fraction of unresolved multiple scatters is estimated to be approximately 5\%-7\% using dedicated simulations and a data-driven approach to predict what fraction of the time multiple scatters merge based on the distribution of S2 time separations. This fraction is included in the fitting routine with a flat prior ranging up to 20\%.

\begin{figure}[!t]
	\centering
	\includegraphics[width= \columnwidth]{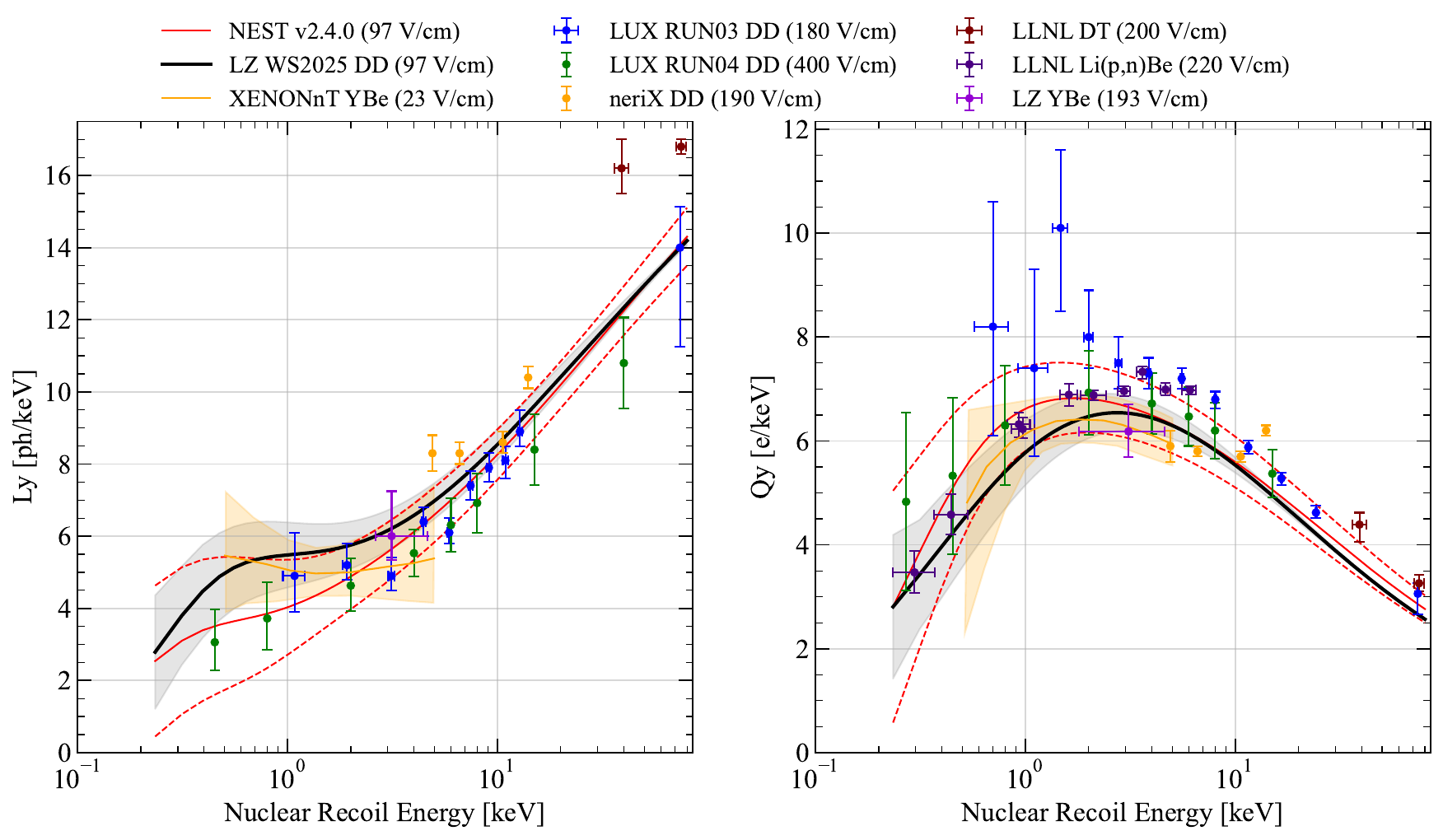}
	\caption{The mean light yield, $L_y$, and charge yield, $Q_y$, used in this analysis are shown as black lines. The $\pm1\sigma$ uncertainty is shown as shaded gray envelope which considers the yield uncertainties from the calibration procedure discussed in the text. The mean and $\pm$1$\sigma$ $Q_y$ and $L_y$ curves from the predicted \textsc{nest}~v2.4.0 NR response model at 97~V/cm is shown in red~\cite{nest_2_4_0}. Yield measurements from LZ~\cite{LZ:2025vdo} and other experimental results are also shown~\cite{XENON:2024kbh,LUX_DD,LUX:2022qxb,nerix:2018,Pershing_2022,      Lenardo:2019fcn,huang}.
    }
	\label{fig:Yields}
\end{figure}

The systematic uncertainties on $L_y$, $Q_y$, and fluctuations in the detector response model are propagated into the statistical inference as discussed in the text. The yield uncertainties are quantified by sampling the MCMC posterior distribution to produce $L_y$ and $Q_y$ curves as a function of energy. These yield curves are then eigen-decomposed, and the two leading order eigenvalues and eigenvectors are mapped back onto corresponding \textsc{nest} parameters to produce the $\pm$1$\sigma$ uncertainty on the best-fit NR yields. The range of yields tested is shown in Fig.~\ref{fig:Yields} and the eigen-decomposition enables us to model both correlated and anti-correlated behavior between $L_y$ and $Q_y$ at threshold. For the fluctuation parameters, only the effect of $F_\text{ex}$ is considered in the final uncertainty calculation. Uncertainties associated with $F_\text{i}$ and recombination fluctuations were determined to have a subdominant impact on the predicted NR signal rate in the ROI. These considerations allow us to reduce the NR response uncertainties on the predicted number of $^{8}$B CE$\nu$NS by approximately 30\% relative to the default \textsc{nest}~v2.4.0 model~\cite{nest_2_4_0}. 

We calculate a combined rate effect of the various systematic uncertainties due to NR modeling and the signal detection efficiencies, parameterized by a single nuisance parameter. This reduces the number of fitted parameters in the statistical inference procedure described in the main text, which would otherwise introduce degeneracies in the fits. We sample from unit Gaussian distributions for each uncertainty and evaluate the product of the corresponding changes in rate to build a combined effective uncertainty. The assumption that the combined rate change can be constructed as a product of the individual rate changes was explicitly confirmed. This procedure is repeated for each NR component -- $^{8}$B CE$\nu$NS, DM, and neutrons. We find that incorporating shape effects due to model uncertainties has a negligible impact on the predicted \{\sonec,~\stwoc\} spectra at the level of predicted statistics in the ROI. We also ignore correlations between individual uncertainties, which is a conservative treatment as including correlations would slightly reduce the overall rate uncertainty. The nuisance parameter, $\sigma_\text{eff}$, is included in the statistical inference which controls the relative rate uncertainty for each NR component. 

\begin{table}[h!]
    \centering
    \caption{Default and best-fit model parameters for the \textsc{nest} NR mean yield (top), \textsc{nest} NR fluctuations response model (middle), and model nuisance parameters (bottom). See Eq.~6 of Ref.~\cite{szydagis2022review} for original definition of yield and fluctuation parameters. In the column labeled \emph{LZ Prior}, the values in the square brackets define upper and lower boundaries of the range in which a parameter may freely float. The values in the curly brackets define the mean and standard deviation of the gaussian providing a constraint within the fit.}
    \begin{tabular}{P{4.2cm} c P{3.2cm} P{3cm} P{3.2cm}}
    \hline
    \hline
    NR Yield Parameter & \textsc{nest} Default & LZ Prior & LZ Posterior \\
    \hline
    $\alpha$ & $11.0^{+2.0}_{-0.5}$ & [8.0, 17.0] & $11.6\pm1.1$ \\
    $\beta$ & 1.1 $\pm$ 0.05 & [0.80, 1.25] & $1.084^{+0.024}_{-0.021}$ \\
    $\gamma$ & $(4.80\pm0.21)\times10^{-2}$ & [0.015, 0.065] & $(5.29^{+0.05}_{-0.06})\times10^{-2}$ \\
    $\epsilon$ [keV] & $12.6^{+3.4}_{-2.9}$ & [2.0, 30.0] & $8.0^{+1.8}_{-1.7}$ \\
    $\zeta$ [keV]  & $(3.0\pm1.0)\times10^{-1}$ & \{0.3, 0.1\} & $(4.4\pm1.5)\times10^{-1}$\\
    $\eta$ & 2 $\pm$ 1&  \{2.0, 1.0\} & $1.1^{+0.5}_{-0.3}$ \\
    $\theta$ [keV] & $(3.0\pm0.5)\times10^{-1}$ & \{0.3, 0.05\} & $(3.0^{+0.8}_{-0.9})\times10^{-1}$ \\
    $\iota$ & 2 $\pm$ 0.5 & \{2.0, 0.5\} & $2.2^{+0.8}_{-0.7}$ \\
    \hline
    \hline 
    \\
    \hline
    \hline
    NR Fluctuations Parameter & \textsc{nest} Default & LZ Prior & LZ Posterior \\
    \hline
    $F_\text{i}$ (constant) & 0.4 & [0.001, 2.000] & $1.1^{+0.3}_{-0.4}$ \\
    $F_\text{ex}$ (constant) & 0.4 & [0.001, 8.000] & $1.1^{+1.4}_{-0.7}$ \\
    $F_\text{ex}$ (linear slope) & - & [0.0, 0.1] & $(2.2\pm0.6)\times10^{-2}$ \\
    $\lambda$ (recombination variance)  & 1 & [0.01, 1.2] & $(2.9^{+2.2}_{-1.7})\times10^{-1}$ \\
    \hline
    \hline
    \\
    \hline
    \hline
    \multicolumn{2}{c}{Nuisance Parameter} & LZ Prior & LZ Posterior \\
    \hline
    \multicolumn{2}{c}{$<2$~MeV neutron fraction} & [0.0, 0.3] & $(1.3\pm0.3)\times10^{-1}$ \\
    \multicolumn{2}{c}{JENDL/(ENDF+JENDL) Fraction} & [0.0, 1.0] & $(7.0^{+1.3}_{-1.4})\times10^{-1}$ \\
    \multicolumn{2}{c}{Unresolved multi-scatter background fraction} & [0.0, 0.2] & $(5.0^{+1.1}_{-1.0})\times10^{-2}$ \\
    \multicolumn{2}{c}{Accidental coincidence background fraction in low energy region} & $0.037$ & - \\
    \hline
    \hline
    \end{tabular}
    \label{tab:nest_params}
\end{table}

\section{Details on Bias Mitigation}
\label{sup:salt}

As described in main text, artificial $^{8}$B CE$\nu$NS-like and DM-like signal events were randomly injected into the data pipeline. The rate and distribution of the DM-like signal events are discussed in detail in the text and supplements of Ref.~\cite{LZ:2024WIMP}. The artificial $^{8}$B CE$\nu$NS events are constructed in the same manner as the artificial DM-like signal events, but have a distribution following the predicted $^{8}$B CE$\nu$NS recoil spectrum.

The injected yearly rate of the $^{8}$B CE$\nu$NS artificial events is drawn from a Poisson distribution of mean 4 events and was randomly introduced into the dataset covering 59\% of the data collection period. We chose the rate to match the Poisson uncertainty on the single scatter rate of $^{8}$B CE$\nu$NS in the fiducial liquid xenon volume using the NR response model and fiducial volume definition of Ref.~\cite{LZ:2024WIMP}. The NR response model described in this paper predicts approximately 1.7 times more $^{8}$B CE$\nu$NS events than assumed in Ref.~\cite{LZ:2024WIMP}. After considering detection efficiencies and live time losses, the expected number of artificial events from both high mass DM and $^{8}$B CE$\nu$NS in our dataset follows a Poisson distribution with a mean of 1.75 events. The number of events injected into the raw data is allowed to be any number (including zero) drawn from this distribution. The true number of injected events is still hidden to the collaboration so as to not impact the bias mitigation of ongoing and future analyses. Only the identity of the 1 event passing the ROI and fiducial volume selection of this analysis was revealed. This event passed all of the data selection criteria described in the main text.

\section{Limits in Terms of Number of DM Events}

\begin{figure}[!t]
	\centering
	\includegraphics[width=0.6 \columnwidth]{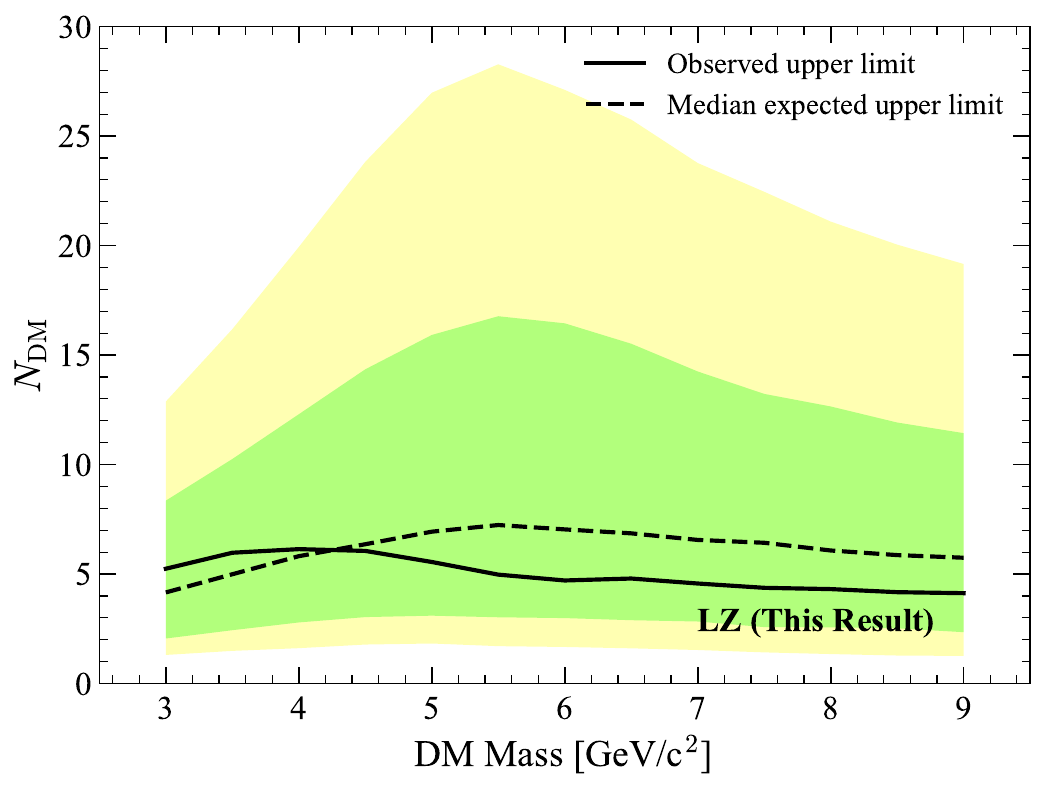}
	\caption{Upper limits (90\% confidence level) on the number of DM events as a function of its mass are shown as a solid black line, for the spin-independent case. Green and yellow regions show the range of expected upper limits from 68\% and 95\% of the background-only experiments, respectively. The dashed black line shows the median expected upper limit using the post-fit model.
    }
	\label{fig:nWIMPs}
\end{figure}

Figure~\ref{fig:nWIMPs} shows the the WS2025 upper limits in terms of number of DM events. 

\end{document}